# Control of Grid-Forming VSCs: A Perspective of Adaptive Fast/Slow Internal Voltage Source

Heng Wu, *Member*, *IEEE*, and Xiongfei Wang, *Fellow*, *IEEE*

*Abstract*—Grid-forming (GFM) capability requirements are increasingly imposed on grid-connected voltage-source converters (VSCs). Under large grid disturbances, GFM-VSCs need to remain stable while providing GFM services. Yet, such objectives, as pointed out in this paper, inherently lead to conflicting requirements on the dynamics of internal voltage source (IVS) of GFM-VSCs, i.e., the fast IVS dynamics is needed to avoid the loss of synchronism with the grid, whereas the slow IVS dynamics is preferred for maintaining GFM capability. To tackle this challenge, an adaptive fast/slow IVS control is proposed, which switches GFM-VSC between fast and slow IVS dynamics based on system needs. The proposed method enhances the transient stability of GFM-VSC, whilst maximizing its capability of providing GFM service. Further, the approach is robust to different grid strengths and different types of grid disturbances. The experimental results verify the theoretical findings and the effectiveness of the proposed control method.

## I. INTRODUCTION

The grid-forming (GFM) control is emerging as a promising approach for massive integration of voltage-source converters (VSCs) into electrical grids [1], [2]. Differing from the legacy grid-following (GFL) control that operates VSC as an *external-voltage-dependent current source*, the GFM-VSC is controlled as a *voltage source behind the impedance*, and hence, its stable operation does not rely on the voltage stiffness of ac grid [3], [4]. It is well documented that GFM-VSCs exhibit superior small-signal stability over GFL-VSCs under weak ac grids [5]-[6]. The recent study further shows that with the proper design of the controller, the small-signal stability of GFM-VSC can be guaranteed with a wide variation of grid strength and operating points [7].

On top of the small-signal stability, control of GFM-VSC under large grid disturbances, including voltage sags, phase-angle jumps, high rate of change of frequency (RoCoF), and the combination thereof, is also of concern. Except for current limitation [8]-[10], GFM-VSCs should also remain synchronism with the grid (i.e., transient stability) during and after different large disturbances [11]. Extensive studies are reported to enhance transient stability of GFM-VSCs, which can, in general, be categorized into two groups:

- Guaranteeing the existence of equilibrium points (EPs) during large disturbances, which is typically realized by modifying the active/reactive power references of GFM-VSC [12]- [14], and by optimizing the current limitation algorithm [15].
- Optimizing the dynamic response of GFM-VSC to guarantee the system convergence to stable EPs (SEPs), which can be achieved by increasing the equivalent damping of GFM-VSC through shaping the active/reactive power control [16]-[19], or using the mode-adaptive control [20]- [21].

While the transient stability is an essential requirement, grid codes are evolving with more capability requirements on GFM-VSCs. More specifically, GFM-VSCs are expected to provide adequate services under grid disturbances, including the provisions of active inertia power to alleviate the RoCoF of the power grid, active phase jump power to counteract the step change of phase angle of grid voltage, as well as the fast fault current injection to boost system voltage during grid faults [22]. Those GFM capabilities are crucial for restoring the power system from abnormal events [23]. Hence, many research efforts have also been made to enhance the GFM capability of VSC through optimizing its active/reactive power control loops [24]-[26].

The state-of-the-art control solutions have made significant progresses in enhancing either transient stability or GFM capabilities of VSCs under large grid disturbances. However, few of them address these two requirements at the same time, even though it is demanded in practice [16]-[18]. In [16], it is demonstrated that increasing virtual inertia may lead to improved GFM capability in alleviating the RoCoF of the system, but at the cost of degraded transient stability. Subsequently, [17]-[18] proposed different damping solutions to mitigate this conflict. However, these solutions have several limitations:
- They have only been tested under specific grid strength and single type of grid disturbance, namely grid voltage dips.
- The entire analysis relies on a strong assumption that the current limit of the GFM-VSC is not triggered during grid disturbances, which might not hold in practice [16]-[18].

To tackle the abovementioned challenges, this paper, thus, revisits the underlying control philosophy that guarantees the transient stability and GFM capability of GFM-VSCs, from the perspective of internal voltage source (IVS) dynamics. *It*

H. Wu is with the AAU Energy, Aalborg University, 9220 Aalborg, Denmark (e-mail: hew@energy.aau.dk).
X. Wang is with KTH Royal Institute of Technology, 10044, Stockholm, Sweden, and also with the AAU Energy, Aalborg University, 9220 Aalborg, Denmark. (e-mail: xiongfei@kth.se). (*Corresponding author: Xiongfei Wang.*)

*is, for the first time, revealed that these two control objectives impose conflicting requirements on the IVS dynamics of GFM-VSC.* More specifically, a slow IVS dynamics is demanded to facilitate GFM-VSC to provide GFM services [4], yet on the other hand, a fast IVS dynamics is needed for enhancing the transient stability of GFM-VSCs. *Therefore, existing control solutions that using the fixed slow IVS control for GFM capability enhancement tend to jeopardize the transient stability of the system, and vice versa.* To tackle this challenge, a high-level, implementation-agnostic control framework named as adaptive fast/slow IVS control, is proposed in this work to better coordinate the requirement of GFM capability and transient stability. The approach furnishes GFM-VSCs with slow IVS control when there are sufficient stability and current margins for providing GFM service, yet operates GFM-VSC with fast IVS control when its current limit is reached or it operates with large power angle, in order to prevent the loss of synchronization (LOS). A detailed realization of such adaptive fast/slow IVS control is further put forward, including:

- A control-mode switching criterion that is based on the local output current magnitude of VSC is proposed, which is readily implemented in practice.
- A standard slow active/reactive power control scheme is adopted for enabling slow IVS dynamics of the GFM-VSC.
- Two widely used control methods for transient stability enhancement, i.e., hybrid-synchronization control [12], [27]-[28], and voltage magnitude based active power refence adjustment [13]-[14],[29], are revisited from the fast IVS control perspective. It is revealed that they have degraded performance under the weak grid. A fast IVS control that modifies the active power reference based on the output current magnitude is thus added to enhance the transient stability of GFM-VSC under the weak grid.

Finally, experimental tests are performed to corroborate the theoretical analysis and the proposed control approach. The proposed adaptive fast/slow IVS control outperforms the state-of-the-art solutions in [16]-[18] with following advantages:

1) Robustness against various grid strengths (from very weak grid to stiff grid) and different types of grid disturbances, including grid voltage dips, phase angle jumps, high RoCoF, and the combinations thereof.
2) The proposed scheme does not rely on any assumptions and remains effective even when the current limit of the GFM-VSC is triggered during grid disturbances.

Moreover, the proposed adaptive fast/slow IVS control is an open framework without any limitations on how fast/slow IVS control should be implemented (although an implementation example is given in this work). Hence, the more advanced fast and/or slow IVS control developed in the future can also be seamlessly incorporate within this control framework for further performance improvement.

## II. IMPACT OF FAST AND SLOW IVS CONTROL

### A. System Description

There are many GFM control schemes available in literatures [3]-[4]. Among others, the GFM-VSC with single-loop voltage magnitude (SLVM) control and adaptive virtual impedance (VI)-based current limitation, as shown in Fig. 1, is adopted as the benchmark in this work [30]-[32]. Nevertheless, the core idea of this work is generic and can be easily extended to GFM-VSCs with other control schemes. As shown in Fig. 1, the GFM-VSC is connected to the point of common coupling (PCC) through an LC filter, and $Z_g$ represents the grid impedance. $P_o/Q_o$ and $v_o/i_o$ denote the output active/reactive power and output voltage/current of VSC, while $v_{inv}$ and $v_g$ are the inverter bridge voltage and grid voltage, respectively.

The GFM-VSC synchronizes with the power grid through the active power control (APC), in which the active power-frequency ($P$-$\omega$) droop control as well as the inertial and damping emulation are embedded [33], as shown in Fig. 1. The output of the APC is the angle reference $\theta_{ref}$ that is used for the $dq$ transformation. $\omega_n$ denotes the nominal angular frequency. $D$ is the P–ω droop coefficient. $K_p$ and $H$ represent the virtual damping and inertia constant, respectively. It is worth noting that the emulated virtual inertia has the same effect as that of a low-pass filter (LPF) [34], and hence, no additional LPF is needed after the active power measurement.

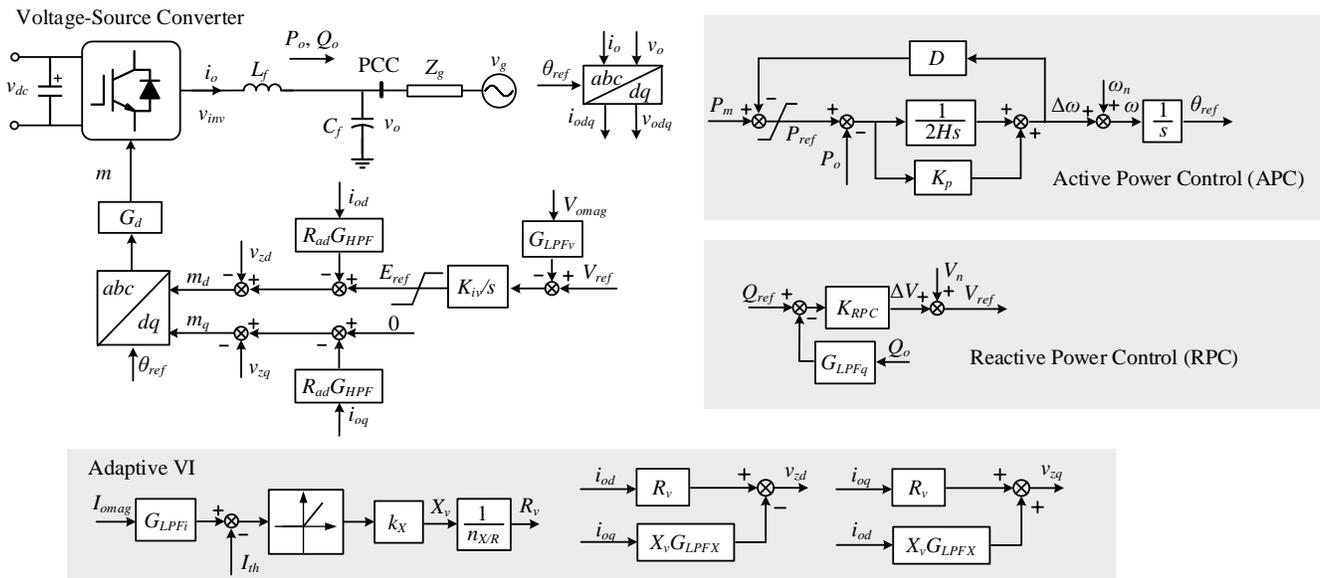

Fig. 1. GFM-VSC with the SLVM control and adaptive VI-based current limitation

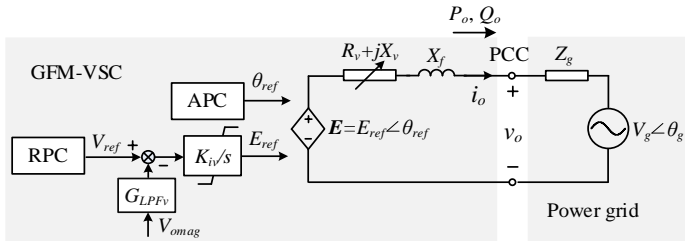

Fig. 2. Equivalent circuit of VSC-grid system.

The voltage magnitude reference of GFM-VSC is regulated by the reactive power control (RPC), e.g., the reactive power-voltage (Q-V) droop control shown in Fig. 1 [3]. $V_{ref}$ is the voltage magnitude reference while $V_n$ is the nominal voltage magnitude. $K_{RPC}$ is the Q–V droop coefficient. $G_{LPFq}$ represents a first-order LPF used for the reactive power measurement.

The SLVM control uses an integrator $K_{iv}/s$ to regulate the output voltage magnitude of VSC ($V_{omag} = \sqrt{v_{od}^2 + v_{oq}^2}$) to track $V_{ref}$ [30], and a first-order LPF $G_{LPFv}$ is adopted to filter out high-frequency noise in $V_{omag}$. The active damping resistor $R_{ad}$ is added at the output of $K_{iv}/s$ to damp the synchronous resonance introduced by the power control [5]. $R_{ad}$ is typically cascaded with a high-pass filter $G_{HPF} = s/(s+\omega_{HPF})$ to eliminate its impact on the steady-state power control performance [5], and $\omega_{HPF}$ represents the cutoff frequency of $G_{HPF}$.

The adaptive VI is used to limit the current of GFM-VSC under large grid disturbances, the general idea of which is to adjust the value of VI based on the fault current magnitude, as shown in Fig. 1. $X_v$ and $R_v$ represent the adaptive virtual reactor and virtual resistor, respectively, which are given by [31]-[32]

$$X_v = \begin{cases} k_X \left( G_{LPFi} I_{omag} - I_{th} \right), & \text{if } I_{omag} \geq I_{th} \\ 0, & \text{if } I_{omag} < I_{th} \end{cases}$$
$$R_v = \frac{X_v}{n_{X/R}} \quad . \quad (1)$$

where $I_{omag}$ denotes the magnitude of output current of VSC, and $I_{th}$ is the threshold current beyond which the adaptive VI will be activated. $k_X$ is the proportional gain of the adaptive virtual reactor and $n_{X/R}$ represents the X/R ratio of the adaptive VI. $G_{LPFi}$ is the first-order LPF used for measuring $I_{omag}$. $v_{zd}$, $v_{zq}$ represent the fictitious voltage drop across the VI, which is calculated as

$$\begin{aligned} v_{zd} &= i_{od} R_v - i_{oq} X_v \\ v_{zq} &= i_{oq} R_v + i_{od} X_v \end{aligned}. \quad (2)$$

The emulation of VI is realized by subtracting the fictitious voltage drop $v_{zd}+jv_{zq}$, from the voltage reference $E_{ref}+j0$, as shown in Fig.1. $G_{LPFX}$ represents the first-order LPF used for calculating $v_{zdq}$, whose purpose is to guarantee the small-signal stability when the adaptive VI is activated [32]. For the detailed parameters tunning of the adaptive VI, please refer to [32].

Fig. 2 illustrates the equivalent circuit of the VSC-grid system, where GFM-VSC is represented by a controlled voltage source behind the impedance. Since the focused timescale of this work is longer than the fundamental period, the filter capacitor branch in Fig.1 can be treated as open circuit due to its large impedance in the low frequency range [5]

### B. Fast or Slow IVS Control: Performance Comparison

This part intends to perform a conceptual comparison of GFM-VSC with slow and fast IVS control. For simplicity, the resistive part of the VI, filter impedance and grid impedance are neglected due to the high X/R ratio in the transmission grid. $X_{tot}$ represents the total reactance, which is calculated as $X_{tot}=X_v+X_f+X_g$. Defining the power angle $\delta$ as the phase difference between the internal voltage phasor of GFM-VSC $E$ and grid voltage phasor $V_g$, i.e., $\delta = \theta_{ref} - \theta_g$, and the output active power of GFM-VSC can be calculated as [Unless otherwise mentioned, the per-unit (p.u.) representation is adopted in this work afterwards] [35]

$$P_o = \frac{E_{ref} V_g}{X_{tot}} \sin \delta \quad . \quad (3)$$

Fig. 3 illustrates the $P_o$-$\delta$ curve as well as phasor diagram of GFM-VSC when subjected to a small grid disturbance, e.g., a small phase angle jump of grid voltage. The solid and dashed lines represent phasors before and after grid disturbances, respectively, while variables after the disturbance are represented with the prime symbol, e.g., $E'$. The GFM-VSC is initially operated at the SEP $a$ with power angle $\delta_s$. By adopting the slow IVS control, the internal voltage phasor $E$ can be kept almost unchanged at the instant of grid phase angle jump, leading to a sudden increase of $\delta$ from $\delta_s$ to $\delta_s'$, and consequent increases of the output current $I_o$ and output active power $P_o$, as shown in Fig. 3(a). The fast provision of this additional active power under phase angle jump of the grid voltage (defined as "active phase jump power" in the grid code [22]) has been proven crucial for maintaining the power system stability. Moreover, since the change of the grid frequency finally ends up with the grid phase angle shifts, the GFM-VSC with a slow IVS control is also capable of providing fast active inertia power injection under grid frequency drifts, which is beneficial for alleviating the RoCoF of the power grid [22].

In contrast, the fast IVS control enables a tight regulation of the output active and reactive power of GFM-VSC to follow their references. This is realized by modifying the internal voltage phasor $E$ in a fast manner to "follow" the change of the grid voltage phasor $V_g$, by which, there is a limit change on $\delta$ (also $I_o$ and $P_o$) under the grid phase angle jump, as shown in Fig. 3(b). In that case, however, the GFM-VSC would have limited responses to grid disturbances, and thus, its benefit in stabilizing the power system is diminished.

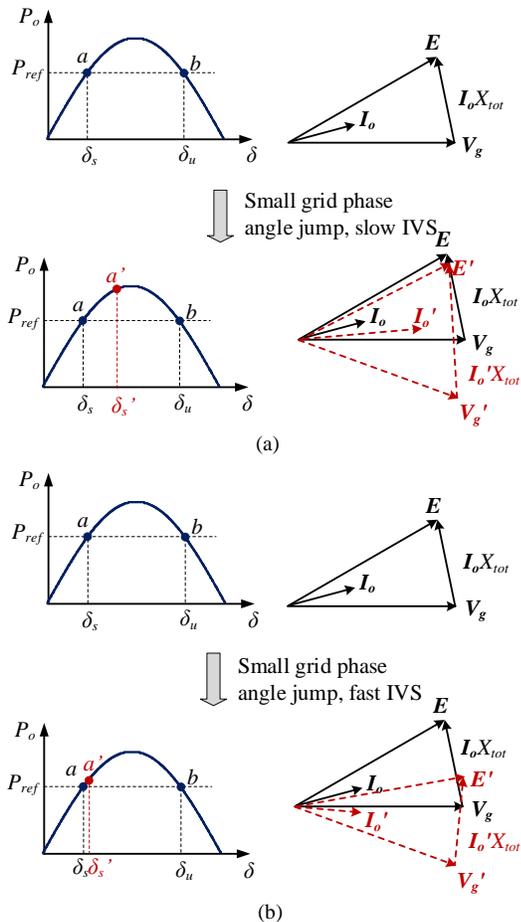

Fig. 3. Phasor diagram of GFM-VSC when subjected to small grid phase jump. The solid and dashed lines represent phasors before and after grid disturbances, respectively, and the prime symbol is used to denote variables after the disturbance (a) With a slow IVS control, additional active output power is generated (b) With a fast IVS control, limited or no additional active output power is generated.

While the slow IVS control is crucial for maintaining GFM capability, it risks in destabilizing GFM-VSC itself under large grid disturbances (e.g., 60° phase angle jump required in the grid code [22]). As shown in Fig. 4 (a), GFM-VSC with slow IVS control might yield a very large $\delta$ under a large phase angle jump, which results in the crossover of the unstable EP (UEP) $b$. Then, since $P_{ref} > P_o$ at new operating point $a'$, $\delta$ will further increase and finally the LOS is resulted [11]. In contrast, the fast IVS control enables the fast modification of internal voltage phasor $E$ to "follow" the change of the grid voltage phasor $V_g$, such that the change of $\delta$ is limited even under a large phase angle jump of the power grid, as shown in Fig. 4(b). Hence, the LOS is avoided.

Table I summarizes the performance comparison of GFM-VSC with fast or slow IVS control, it can be seen that the slow IVS control is beneficial for maintaining GFM capability, yet it can jeopardize the transient stability of VSC under severe grid disturbances. The fast IVS control introduces the opposite effect.

It is worth mentioning that even though the performance comparison of GFM-VSC with fast/slow IVS control is made by considering phase angle jump of the grid voltage, same conclusions also hold for grid voltage dips (grid faults), which can be easily demonstrated by following the similar analysis procedure and will not be repeated here due to the page limit.

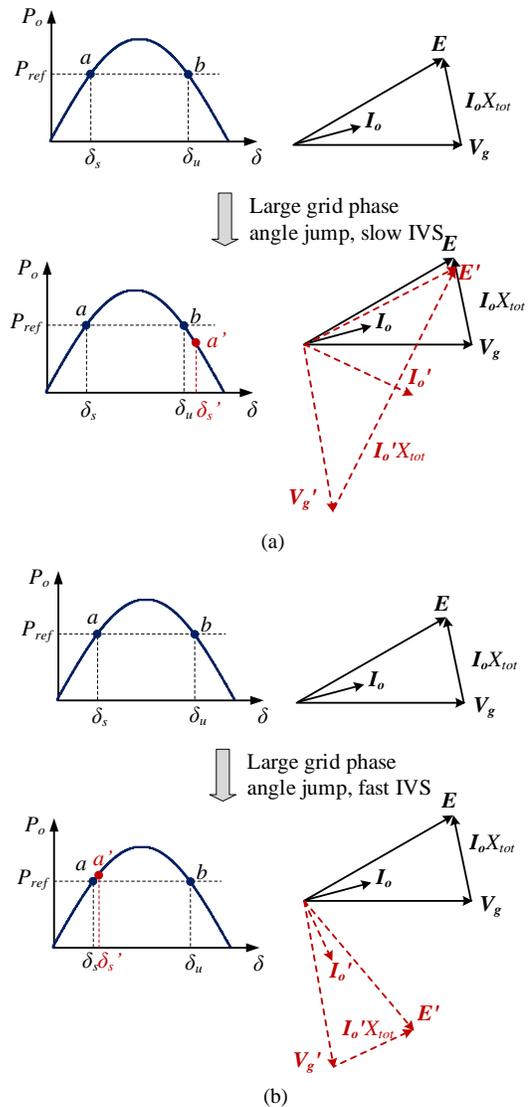

Fig. 4. Phasor diagram of GFM-VSC when subjected to large grid phase jump. The solid and dashed lines represent phasors before and after grid disturbances, respectively, and the prime symbol is used to denote variables after the disturbance. (a) With a slow IVS control, LOS (b) With a fast IVS control, LOS is prevented.

TABLE I
PERFORMANCE COMPARISON BETWEEN FAST AND SLOW IVS CONTROL

|  | **Slow IVS control** | **Fast IVS control** |
|---|---|---|
| Features | Slow dynamics of internal voltage phasor $E$ | Fast dynamics of internal voltage phasor $E$ |
| GFM capability | √ | × |
| Large-signal stability | × | √ |

## III. ADAPTIVE FAST/SLOW IVS CONTROL

### A. Basic Idea

As pointed out in Section II-B, the GFM-VSC with a slow IVS control can naturally provide GFM service to counteract grid disturbances. Yet, from the transient stability perspective, a fast IVS control is preferred to minimize the risk of LOS. These conflicting control requirements make it difficult to simultaneously enhance the transient stability and GFM capability with a fixed control scheme. Therefore, an adaptive fast/slow IVS control is proposed in this work, which switches GFM-VSC between fast and slow IVS control based on its operation conditions, as shown in Fig. 5. The design principles of the adaptive fast/slow IVS control are as follows:

- Since GFM-VSC cannot provide any additional power/current for GFM service after reaching its current limit ($I_{omag} = I_{lim}$), adopting the slow IVS control in this scenario does not make sense. Therefore, GFM-VSC is always operated with the fast IVS control to avoid the LOS under the current limiting mode.
- In the case that the current limit of GFM-VSC is not reached ($I_{omag} < I_{lim}$), the risk of LOS is low when it is operated with small power angle [3]. Consequently, maintaining GFM capability should be prioritized by adopting the slow IVS control. However, the fast IVS control should be used to prevent the LOS when GFM-VSC is operated with large power angle. Hence, the mode switching between fast/slow IVS control can be realized by comparing the power angle $\delta$ with a pre-defined threshold $\delta_{th}$, as shown in Fig. 5. The selection of $\delta_{th}$ will be discussed afterwards.

Nevertheless, it is known from (3) that the calculation of $\delta$ requires the knowledge of grid voltage and of grid impedance [36], which is challenging in practice. Therefore, a switching criterion based on local measurements should be developed for practical implementation. The selected local variable $x$ should hold a monotonic relationship with $\delta$, such that the original switching criterion based on the comparison between $\delta$ and $\delta_{th}$ can be mapped to comparing $x$ and $x_{th}$. Moreover, the impact of different SCR and $V_g$ should be also considered to guarantee the robustness of the switching criterion.

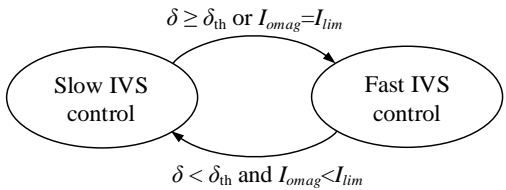

Fig. 5. General idea of the adaptive fast/slow IVS control.

### B. $I_{omag}$-based Switching Criterion

It is known from Fig. 5 that the local measurement variable $I_{omag}$ is always needed to check whether GFM-VSC is operated under the current limit mode. Therefore, it is interesting to first investigate whether $I_{omag}$ holds a monotonic relationship with $\delta$ before reaching the current limitation, such that the $\delta$-based switching criterion can be mapped to the $I_{omag}$-based switching criterion

It is known from Fig. 2 that the internal voltage phasor $E$ is aligned with the $dq$-frame defined by the output phase angle of the APC. Based on Fig. 2, the dq components as well as the magnitude of the output current can be calculated as

$$I_{od} + jI_{oq} = \frac{E_{ref} + j0 - (V_{gd} + jV_{gq})}{jX_{tot}}. \quad (4)$$

$$I_{omag} = \sqrt{I_{od}^2 + I_{oq}^2}. \quad (5)$$

where

$$V_{gd} = V_g \cos\delta. \quad (6)$$

$$V_{gq} = -V_g \sin\delta. \quad (7)$$

$$X_{tot} = X_f + X_g. \quad (8)$$

Substituting (4), (6)-(8) into (5), which yields

$$I_{omag} = \frac{\sqrt{E_{ref}^2 + V_g^2 - 2E_{ref}V_g \cos\delta}}{X_f + X_g}. \quad (9)$$

Substituting $\delta = \delta_{th}$ into (10), which yields

$$I_{omag\_th} = \frac{\sqrt{E_{ref}^2 + V_g^2 - 2E_{ref}V_g \cos\delta_{th}}}{X_f + X_g}. \quad (10)$$

It is known from (9) that $I_{omag}$ is monotonously increased with the increase of $\delta$ before triggering the current limitation, which yields $I_{omag} < I_{omag\_th} \leftrightarrow \delta < \delta_{th}$. Yet, it is clear from (10) that $I_{omag\_th}$ is not a constant value for specific $\delta_{th}$, but is affected by two other variables, i.e., grid impedance $X_g$ and grid voltage $V_g$. Define $I_{omag\_th\_min}$ is the minimum value of $I_{omag\_th}$ considering different $X_g$ and $V_g$, the sufficient condition for $\delta < \delta_{th}$ becomes $I_{omag} < I_{omag\_th\_min}$. To calculate $I_{omag\_th\_min}$, the largest possible value of $X_g$, i.e., $X_g=1$p.u. (corresponding SCR=1) is considered, as $X_g$ appears in the denominator of (10). On the other hand, the value of $V_g$ for calculating $I_{omag\_th\_min}$ can be obtained by solving

$$\frac{dI_{omag\_th}}{dV_g} = \frac{d\left(\sqrt{E_{ref}^2 + V_g^2 - 2E_{ref}V_g \cos\delta_{th}}\right)}{dV_g} = 0. \quad (11)$$

which yields

$$V_g = E_{ref} \cos\delta_{th}. \quad (12)$$

Since $V_g$ cannot be a negative value, the value of $V_g$ that is used for calculating $I_{omag\_th\_min}$ is given by

$$V_g = \begin{cases} E_{ref} \cos\delta_{th}, & \text{if } \delta_{th} < 90° \\ 0, & \text{if } \delta_{th} \geq 90° \end{cases}. \quad (13)$$

Substituting (13) into (10), $I_{omag\_th\_min}$ can be calculated as

$$I_{omag\_th\_\min} = \begin{cases} \dfrac{E_{ref}\sqrt{1-\cos^2\delta_{th}}}{X_f + X_g}, & \text{if } \delta_{th} < 90° \\ \dfrac{E_{ref}}{X_f + X_g}, & \text{if } \delta_{th} \geq 90° \end{cases}. \quad (14)$$

Considering a typical 0.1 p.u. filter reactance, i.e., $X_f = 0.1$ p.u., together with $X_g = 1$p.u. and $E_{ref} = 1.1$p.u. (In order to maintain 1p.u. PCC voltage, $E_{ref}$ is slightly higher than 1p.u. to compensate the voltage drop across $X_f$ [30]), the $I_{omag\_th\_\min} - \delta_{th}$ curve can be plotted based on (14), as shown in Fig. 6. It is clear that the maximum value of $I_{omag\_th\_\min}$ is around 1 p.u., which is lower than current limit of GFM-VSC, e.g., $I_{lim} = 1.5$ p.u. selected in this work. Therefore, $I_{omag} < I_{omag\_th\_\min}$ is the sufficient condition for both $\delta < \delta_{th}$ and $I_{omag} < I_{lim}$. Therefore, two switching criteria given by Fig. 5 can be combined, i.e., the GFM-VSC is operated with a slow IVS control provided $I_{omag} < I_{omag\_th\_\min}$, otherwise it is switched to the fast IVS control.

The value of $I_{omag\_th\_\min}$ is dependent on $\delta_{th}$, which should be selected smaller than $\delta_u$ (i.e., the corresponding power angle of the UEP *b*, see Fig. 4) to avoid the LOS [35]. It is noted that $\delta_u = 90°$ is yielded under the worst case where SCR=1 [3], and hence, $\delta_{th} < 90°$ is required. A smaller $\delta_{th}$ can better guarantee the transient stability of GFM-VSC, yet it jeopardizes the GFM capability. Therefore, a tradeoff is needed in practice. In this paper, $\delta_{th} = 70°$ selected, which leads to $I_{omag\_th\_\min} = 0.94$ p.u. based on (14).

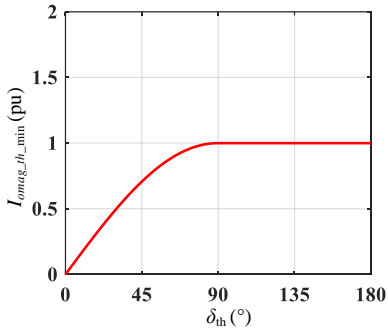

Fig. 6. $I_{omag\_th\_\min} - \delta_{th}$ curve.

### C. Discussion and Practical Realization

It should be noted that $I_{omag} = 0.94$ p.u. ↔ $\delta = 70°$ only holds under the worst scenario where SCR=1 and $V_g$ satisfies (13). In other cases, e.g., stiffer grid conditions, $I_{omag}$ may reach 0.94 p.u. with a much smaller $\delta$. As seen from the $I_{omag} - \delta$ curve given by Fig. 7, $I_{omag} = 0.94$ p.u. ↔ $\delta = 10°$ is yielded with SCR=10 and $V_g=1$p.u. In this scenario, one might argue that the proposed $I_{omag}$-based switching criterion may unnecessarily switch GFM-VSC to the fast IVS control when it is operated with a small $\delta$ (e.g, 10°). However, it should be noted that the current limitation of GFM-VSC will also be triggered with a small $\delta$ under the stiff grid [$I_{omag} = I_{lim}$ (1.5 p.u.) ↔ $\delta = 15°$ when SCR=10, see Fig. 7]. Therefore, even if the ideal $\delta$-based switching criterion in Fig. 5 can be used, the GFM-VSC would still be switched to the fast IVS control with a small $\delta$ (e.g, 15°), due to the triggering of the current limitation. Hence, it is true that the worst-case based design of $I_{omag}$-based switching criterion may jeopardize the GFM capability of VSC under the stiff grid compared with the ideal $\delta$-based switching criterion, the degradation is not significant considering the presence of the current limit control.

Fig. 8 illustrates the practical switching criterion used in this work. A hysteresis comparator and a time delay are added to avoid the frequent switching of control modes caused by the measurement noise. The GFM-VSC is switched to the fast IVS control when $I_{omag} > I_{omag\_th\_\min}$, but can only be switched back to the slow IVS control if $I_{omag} \leq 0.9 I_{omag\_th\_\min}$ holds for a given amount of time ($t_1$). A larger $t_1$ leads to a better noise immunity of the criterion but it degrades the sensitivity. A compromise is, thus, needed in the practical application. In this work, $t_1 = 0.2$s is adopted.

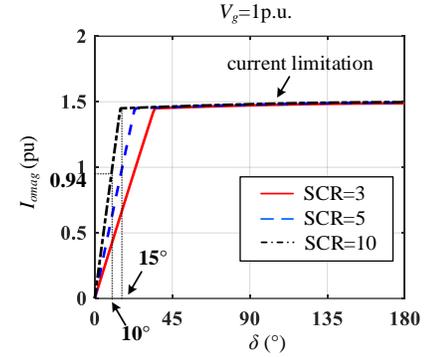

Fig. 7. $I_{omag}$-$\delta$ curve of GFM-VSC. $V_g=1$p.u. with different SCR.

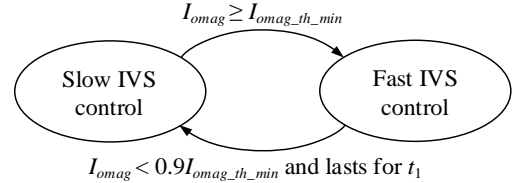

Fig. 8. Practical switching criterion of the proposed adaptive fast/slow IVS control.

## IV. FAST IVS CONTROL IMPLEMENTATION

### A. Basic Idea

Following the practical switching criterion given in Fig. 8, the detailed implementation of the adaptive slow/fast IVS control should be further investigated. Due to the low control bandwidth of the APC/RPC, the GFM-VSC is inherently a slow IVS when it is operated with the standard APC/RPC given in Fig. 1 [4]. Therefore, only modified control schemes that furnish GFM-VSC with a fast IVS will be discussed in this section. It is worth noting that even though the modifications can be made on both APC/RPC, the modification on RPC is not that effective, as RPC is often bypassed under grid faults, due to the saturation of $E_{ref}$ in Fig. 1. Therefore, the implementation of the fast IVS control would only consider the modification of APC in this work.

The underlying idea of fast IVS control is to retain the power angle $\delta$ under grid disturbances to prevent the LOS. This can be realized by modifying $\theta_{ref}$ in the APC to guarantee that $\theta_{ref}$ could follow $\theta_g$ under grid disturbances. In the following, two existing solutions that modify $\theta_{ref}$ through $\omega$ and $P_{ref}$ of APC are re-visited from the perspective of the fast IVS control, and their effectiveness under different grid strengths/disturbances are discussed. To circumvent the drawbacks of the two existing methods, a new fast IVS control that modifies $P_{ref}$ based on the $I_{omag}$ is added.

### B. Hybrid-Synchronization Control

Fig. 9 illustrates the block diagram of the hybrid-synchronization control (HSC) [12],[27]-[28], in which $V_{oq}$ is feedforwarded to modify $\omega$ through a proportional gain $K_{pvq}$. It is pointed out in [12] and [27] that the transient stability of GFM-VSC can be improved by adopting the HSC, whose underlying mechanism is re-visited here from the perspective of the fast IVS control. For a better illustration, the detailed expression of $V_{oq}$ will be derived first.

It is known from Fig. 2 that

$$V_{od} + jV_{oq} = V_{gd} + jV_{gq} + jX_g\left(I_{od} + jI_{oq}\right). \quad (15)$$

Substituting (4), (6)-(8) into (15), which yields

$$V_{oq} = -(1-a)V_g \sin\delta. \quad (16)$$

where

$$a = \frac{X_g}{X_{tot}}. \quad (17)$$

Therefore, $\Delta\omega_q$ in Fig. 9 can be expressed as

$$\Delta\omega_q = K_{pvq}V_{oq} = -K_{pvq}(1-a)V_g \sin\delta. \quad (18)$$

Based on (18), the mechanism of the HSC that operates GFM-VSC as a fast IVS to prevent the LOS can be easily understood. For any disturbances that cause an increase of $\delta$, a more negative $\Delta\omega_q$ is generated via the $V_{oq}$ feedforwarded path [see (18)], which reduces the $\omega(\theta_{ref})$ of GFM-VSC, and hence, counteracts the effect of $\delta$ increase. Yet, with a fixed $K_{pvq}$, it is also known from (18) that the effect of the $V_{oq}$ feedforward is dependent on two factors: (1-$a$) and $V_g$. More specifically:

- $V_{oq}$ feedforward is more effective under the grid phase angle jump with $V_g$ =1p.u., yet it becomes less effective under grid voltage dip due to the smaller $V_g$.
- $V_{oq}$ feedforward is more effective under the stiff grid where $a$ is close to 0, yet it becomes less effective under the weak grid where $a$ approaches to 1.

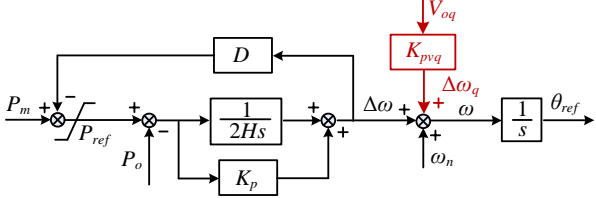

Fig. 9. Block-diagram of the hybrid-synchronization control.

### C. $V_{omag}$-based $P_{ref}$ Adjustment

Another widely used method of avoiding the LOS is to reduce $P_{ref}$ based on the dip of voltage magnitude $V_{omag}$. To that end, different implementations are given in [13]-[14], [29].

Among others, the simplest implementation, i.e., $P_{ref} = P_{ref0}V_{omag}$, is adopted in this work for illustration. The corresponding control diagram is given in Fig. 10, where $V_{omag}$ can be calculated as

$$V_{omag} = \sqrt{V_{od}^2 + V_{oq}^2}. \quad (19)$$

Substituting (4), (6)-(8), (15) into (19), which yields

$$V_{omag} = \sqrt{(1-a)^2 V_g^2 + a^2 E_{ref}^2 + 2a(1-a)E_{ref}V_g \cos\delta}. \quad (20)$$

It is clear from (20) that $V_{omag}$ is reduced with the decrease of $V_g$ and with the increase of $\delta$. Hence, the $V_{omag}$-based $P_{ref}$ adjustment essentially operates the GFM-VSC as a fast IVS, which reduces $P_{ref1}$ under grid faults or grid phase angle jump. The lower value of $P_{ref1}$ helps to reduce $\omega(\theta_{ref})$ of GFM-VSC, which alleviates the risk of LOS. Nevertheless, it can be seen from (20) that the performance of this approach is affected by the grid strength that is characterized by the factor $a$. More specifically:

- $V_{omag} \approx V_g$ under the stiff grid where $a\to 0$, which is independent of $\delta$. Thus, $V_{omag}$-based $P_{ref}$ adjustment is very effective under the grid voltage dip, but less effective under grid phase angle jump.
- $V_{omag} \approx E_{ref}$ under the ultra weak grid where $a\to 1$, which is independent of both $V_g$ and $\delta$. In this scenario, $V_{omag}$-based $P_{ref}$ adjustment is ineffective under both grid voltage dip and grid phase angle jump.

Table II summarizes the performances of HSC and $V_{omag}$-based $P_{ref}$ adjustment in respect to the enhancement of transient stability, considering different grid disturbances and grid strengths. The HSC is more effective under the phase angle jump while $V_{omag}$-based $P_{ref}$ adjustment is more effective under the voltage dip. Therefore, these two methods are complementary to each other, and thus, can be used at the same time for the performance improvement. Nevertheless, the effectiveness of both methods is degraded under weak grid conditions. This is because both of them manipulate the APC based on the PCC voltage, which deviates significantly from the grid voltage under the weak grid. Therefore, a new fast IVS control that can improve the transient stability of GFM-VSC under weak grid conditions is needed.

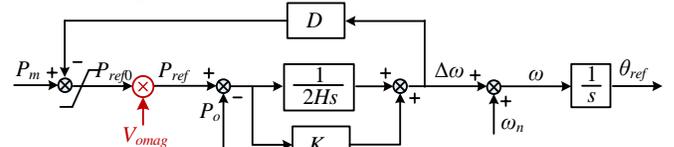

Fig. 10. Control diagram that modifies $P_{ref}$ based on $V_{omag}$

TABLE II
PERFORMANCE OF HSC AND $V_{OMAG}$-BASED $P_{REF}$ ADJUSTMENT

|  | Grid voltage dips |  | Grid phase angle jump/RoCoF |  |
|---|---|---|---|---|
|  | Stiff grid | Weak grid | Stiff grid | Weak grid |
| HSC | Moderate | Low | High | Low |
| $V_{omag}$-based $P_{ref}$ adjustment | High | Low | Low | Low |

### D. $P_{ref}$-$I_{omag}$ Droop Control

It can be seen from (9) that for any given grid strength ($X_g$), $I_{omag}$ is always increased with the increase of $\delta$ [as the term $\cos\delta$ in (9) is decreased with the increase of $\delta$]. Therefore, reducing $P_{ref}$ based on the increase of $I_{omag}$ (which reflects the increase of $\delta$) is a more robust fast IVS control against grid strength variations, which is named as $P_{ref}$ - $I_{omag}$ droop control hereafter. Fig.11 illustrates the control diagram of the proposed method, whose principle can be elaborated as

$$P_{ref1} = \begin{cases} P_{ref}, & \text{if } I_{omag} < I_{Pth} \\ P_{ref} - n(I_{omag} - I_{Pth}), & \text{if } I_{omag} \geq I_{Pth} \end{cases}. \quad (21)$$

where $n$ is the droop coefficient, and $I_{Pth}$ is the threshold current beyond which the proposed $P_{ref}$ - $I_{omag}$ droop control is activated. In this work, $I_{Pth}$ =1.1p.u. is selected to avoid the impact of the proposed method on the steady-state operation of the system.

A larger droop coefficient $n$ indicates more $P_{ref}$ reduction with the same increasement of $I_{omag}$, which is beneficial for system transient stability. Yet, the risk of unnecessary $P_{ref}$ swing due to the measurement noise is also increased. In this work, $n$=10 p.u. is selected for a good compromise, which indicates that a 0.1 p.u. increase of $I_{omag}$ leads to a 1p.u. decrease of $P_{ref}$ when $I_{omag} \geq I_{Pth}$. Moreover, to avoid that $P_{ref1}$ becomes a negative value due to the $P_{ref}$ - $I_{omag}$ droop control, a limiter is added to limit $P_{ref1}$ within 0 to 1 p.u.

Fig. 12 illustrates the final implementation of the proposed adaptive fast/slow IVS control. The fast IVS control is implemented by adding the $P_{ref}$ - $I_{omag}$ droop control, $V_{omag}$-based $P_{ref}$ adjustment, and $V_{oq}$ feedforward on top of the slow IVS control, and is activated/de-activated based on the switching criterion defined in Fig. 8. When the fast IVS control is de-activated, Fig. 12 reduces to the standard control scheme given in Fig. 1, which is a slow IVS control that guarantees the GFM capability of VSC.

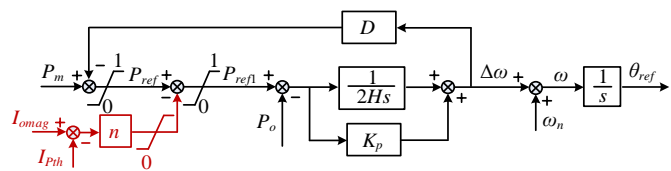

Fig. 11. Block diagram that proposed $P_{ref}$-$I_{omag}$ droop control.

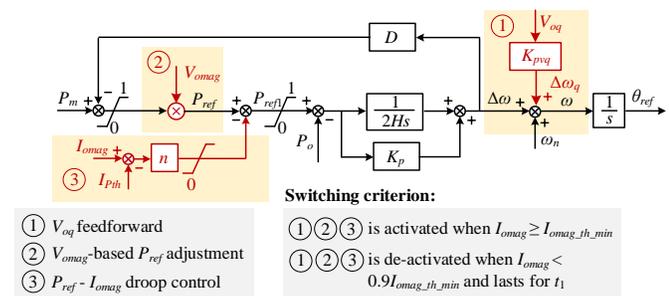

Fig. 12. Detailed implementation of the proposed adaptive fast/slow IVS control.

### E. Small-Signal Stability Analysis

As mentioned, the GFM-VSC is equipped with standard APC/RPC when operated as a slow IVS, whose small-signal stability has been extensively investigated [7],[32],[37]. Hence, the small-signal stability analysis here will focus on the scenario where GFM-VSC is subjected to large grid disturbances with fast IVS control, i.e., $V_{oq}$ feedforward, $V_{omag}$-based $P_{ref}$ adjustment, as well as $P_{ref}$-$I_{omag}$ droop control, activated. The complete small-signal model of GFM-VSC with fast IVS control is derived in the Appendix, based on which, the closed-loop poles of the system can be calculated from (A.21), as shown in Fig. 13. It can be observed that all poles are located in the left half plane, indicating that the system is small-signal stable with used controller parameters (see Table IV) in this work.

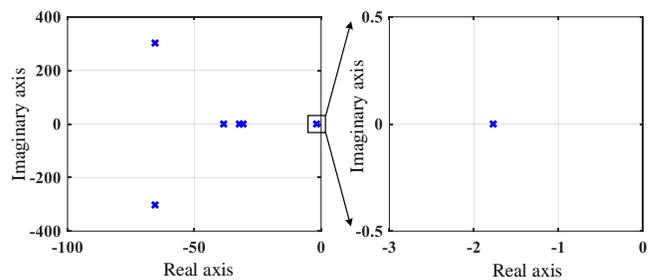

Fig. 13. Pole maps GFM-VSC with fast IVS control activated.

## V. EXPERIMENTAL RESULTS

Experimental tests are carried out to validate the proposed adaptive fast/slow IVS control for the GFM-VSC, and the parameters given in Table III and IV are adopted. The controller parameters in Table IV are tunned based on [7],[32],[37]-[38] to guarantee the small-signal stability of the system. Fig. 14 illustrates the configuration of the experimental setup, where a 45 kVA Chroma 61850 grid simulator is used to generate the grid voltage and emulate different grid disturbances. The control algorithm is implemented in the DS1007 dSPACE system. The DS2102 highspeed D/A board is used to transmit the calculated active power, power angle, current magnitude ($I_{omag}$) and the enable signal of fast IVS control (FastIVS_en) to the oscilloscope.

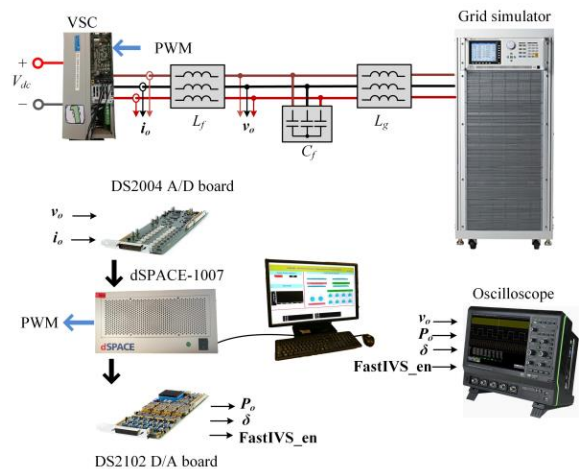

Fig. 14. Configuration of the experimental setup.

TABLE III
SYSTEM PARAMETERS

| SYMBOL | DESCRIPTION | VALUE |
|---|---|---|
| $V_{PCC}$ | PCC voltage (RMS value) | 50 V |
| $P$ | Power rating of the VSC | 1 kW |
| $f_g$ | Grid frequency | 50 Hz |
| $L_f$ | Filter inductance | 3 mH. |
| $C_f$ | Filter capacitance | 20 $\mu$F |
| SCR | Short circuit ratio | Weak grid: 1.2 Stiff grid: 10 |
| $f_{sw}$ | Switching frequency | 10 kHz |
| $T_s$ | Sampling (control) period | 100 $\mu$s |
| $T_d$ | Time delay in the control loop | 1.5$T_s$ |

TABLE IV
CONTROLLER PARAMETERS

| SYMBOL | DESCRIPTION | VALUE (P.U.) |
|---|---|---|
| $n_{X/R}$ | X/R ratio of the adaptive VI | 5 |
| $k_X$ | Proportional gain of adaptive virtual reactance | 1.45 p.u. |
| $I_{th}$ | Threshold current of activating adaptive VI | 1.1 p.u. |
| $I_{lim}$ | Current limit | 1.5 p.u. |
| $K_{iv}$ | Integral gain of SLVM control | 6.28 p.u. |
| $\omega_{LPFX}$ | Cutoff frequency of $G_{LPFX}$. | 0.2 p.u. |
| $R_{ad}$ | Active damping resistor | 0.1 p.u. |
| $\omega_{HPF}$ | Cutoff frequency of $G_{HPF}$. | 0.1 p.u. |
| $D$ | P-$\omega$ droop control | 50 p.u. |
| $K_p$ | Proportional gain of APC | 0.02 p.u. |
| $H$ | Inertial time constant of APC | 10 p.u. |
| $K_{RPC}$ | Q-V droop control | 0.1 p.u. |
| $\omega_q$ | Cutoff frequency of $G_{LPFq}$ | 1 p.u. |
| $I_{omag\_th\_min}$ | Threshold current for fast/slow IVS control switching | 0.94 p.u. |
| $K_q$ | Proportional gain of HSC | 0.34 p.u. |
| $I_{Pth}$ | Threshold current of activating $P_{ref}$ - $I_{omag}$ droop control | 1.1 p.u. |
| $n$ | Droop coefficient of $P_{ref}$ - $I_{omag}$ droop control | 10 p.u. |

### A. Verification of Switching Criterion

Experimental tests are carried out to verify the $I_{omag}$-based switching criterion given in Fig. 8. As shown in Fig. 15, the GFM-VSC is initially operated with slow IVS control (FastIVS_en=0) when $I_{omag}$ ($P_m$ =0.1 p.u.) is less than the minimum threshold, i.e., $I_{omag} < I_{omag\_th\_min}$ = 0.94 p.u. When a symmetrical fault occurs with $V_g$ reduced to 0.1 p.u., the GFM-VSC is operated under the current limiting mode where $I_{omag}$ exceeds the minimum threshold, i.e., $I_{omag}$ = 1.5 p.u > $I_{omag\_th\_min}$, and the fast IVS control is thus activated (FastIVS_en=1). After the fault is cleared, the $I_{omag}$ is reduced to be less than 0.9$I_{omag\_th\_min}$, yet the GFM-VSC is still operated with the fast IVS control, as expected, for 0.2 s before it switches back to the slow IVS control. The experimental results given in Fig. 15 matches well with the proposed switching criterion given in Fig. 8.

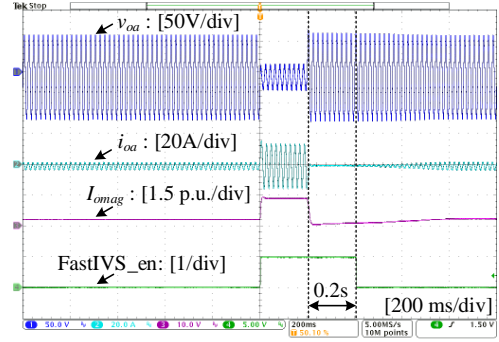

Fig.15. Dynamic response of GFM-VSC with adaptive fast/slow IVS control during grid faults ($V_g$ drops to 0.1 p.u. and lasts 0.2s). SCR=10, $P_m$=0.1 p.u.

### B. Verification of GFM Capability

In order to verify the GFM capability of VSC with the proposed adaptive fast/slow IVS control, experimental tests are carried out by operating GFM-VSC under the stiff grid condition (SCR=10) with zero output power ($P_m$ =0 p.u.) before grid disturbances. For better illustration, two types of control schemes are tested and compared:

1. Adaptive fast/slow IVS control: corresponds to the control scheme given in Fig. 12.
2. Fast IVS control: corresponds to the control scheme given in Fig. 12 with $P_{ref}$ - $I_{omag}$ droop control, $V_{omag}$-based $P_{ref}$ adjustment, and $V_{oq}$ feedforward always enabled.

Fig. 16 shows the dynamic responses of GFM-VSC when subjected to -5Hz/s RoCoF event of the power grid that lasts 0.1s. Since GFM-VSC is operated with small power angle before and after grid disturbances without triggering the current limitation, it would always operate with the slow IVS control (FastIVS_en=0) when the proposed adaptive control is adopted. Hence, despite of generating 0.5 p.u. droop-control-induced active power due to 0.5Hz frequency dips, the GFM-VSC is also capable of providing additional active inertia power to counteract the RoCoF, as shown in Fig. 16(a). For comparison, Fig. 16(b) illustrates the dynamics of GFM-VSC with fast IVS control always enabled (FastIVS_en=1), it can be seen that the capability of GFM-VSC in providing active inertial power is significantly degraded.

Fig. 17 shows the dynamic responses of GFM-VSC when subjected to -10° phase angle jump of the power grid ($\theta_g$= -10°), and GFM-VSC is operated with small power angle before and after grid disturbances without triggering the current limitation. By adopting the proposed adaptive control, GFM-VSC would remain operating with slow IVS control (FastIVS_en=0), which maintains its capability of providing active phase jump power, as shown in Fig. 17(a). In contrast, it can be observed from Fig. 17(b) that much lower active phase jump power is generated from GFM-VSC under same grid phase angle jump, provided that GFM-VSC is always operated with the fast IVS control (FastIVS_en=1). The experimental results in Figs. 16-17 clearly demonstrate the benefit of the proposed fast/slow IVS control in retaining the GFM capability of VSC.

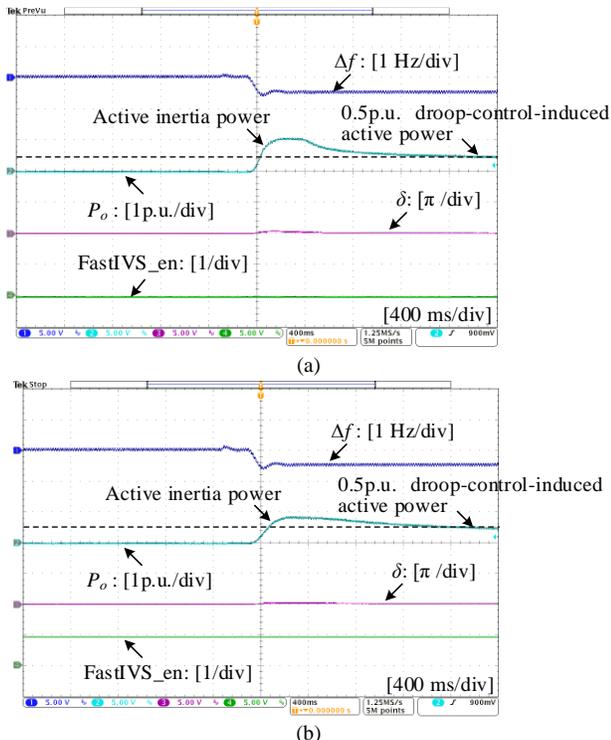

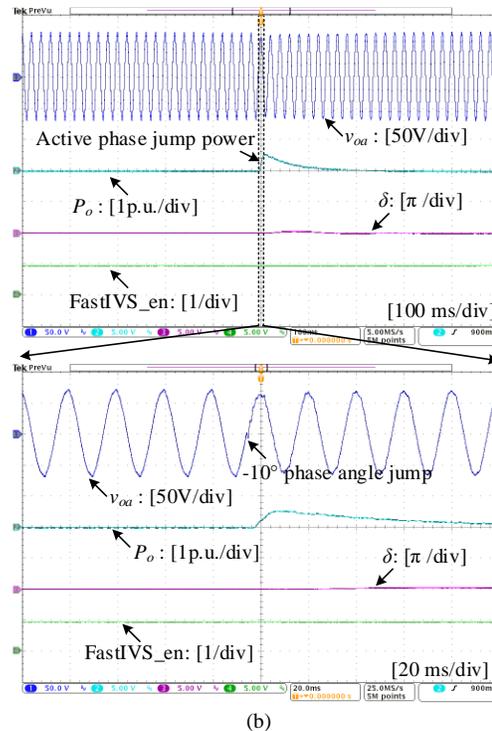

Fig. 16. Dynamic response of GFM-VSC when subjected to -5Hz/s RoCoF that lasts 0.1s, SCR=10, $P_m$=0p.u. (a) Adaptive fast/slow IVS control. (b) Fast IVS control.

Fig. 17. Dynamic response of GFM-VSC when subjected to -10° grid phase angle jump ($\theta_g$= -10°). SCR=10, $P_m$=0p.u. (a) Adaptive fast/slow IVS control. (b) Fast IVS control

## C. Verification of Transient-Stability Improvement

In this part, more severe grid disturbances are considered to verify the transient stability performance of GFM-VSC with the proposed adaptive fast/slow IVS control. For better illustration, three types of control schemes are tested and compared:
1. Slow IVS control: corresponds to the basic GFM control scheme given in Fig. 1.
2. Adaptive fast/slow IVS control without $P_{ref}$-$I_{omag}$ droop: corresponds to the control scheme given in Fig. 12, yet the $P_{ref}$ - $I_{omag}$ droop control is not added.
3. Adaptive fast/slow IVS control: corresponds to the control scheme given in Fig. 12.

*1) Stiff grid (SCR=10)*

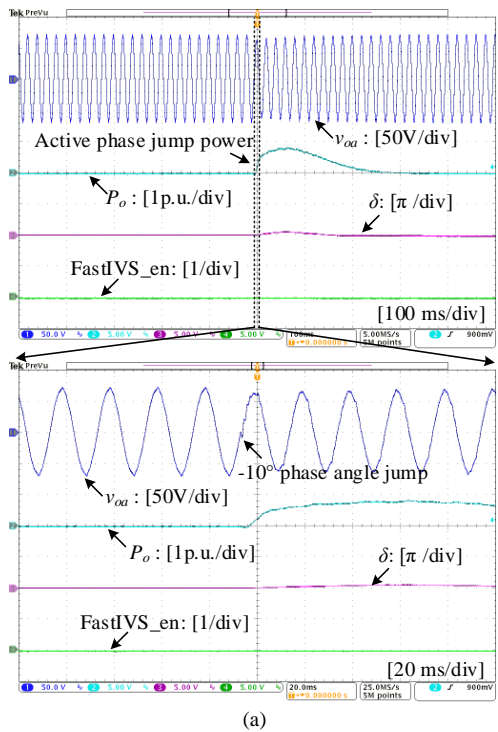

(a)

Fig. 18 illustrates the dynamic responses of GFM-VSC under the stiff grid condition (SCR=10) considering phase angle jump+RoCoF disturbances [-60° phase angle jump ($\theta_g$= -60°) together with -5Hz/s RoCoF that lasts 0.2s]. The GFM-VSC is initially operated with $P_m$ =0.4 p.u. It can be seen from Fig. 18(a) that GFM-VSC becomes unstable by adopting the slow IVS control (FastIVS_en=0), where the oscillation in the output power and power angle can be clearly observed, indicating the LOS of GFM-VSC. In contrast, GFM-VSC can successfully ride through large phase angle jump+RoCoF disturbances by using the adaptive control with or without $P_{ref}$ - $I_{omag}$ droop. As illustrated in Figs. 18(b) and (c), the adaptive control initially operates GFM-VSC with the slow IVS control (FastIVS_en=0), but is naturally switched to the fast IVS control (FastIVS_en=1) under grid disturbances to facilitate the grid synchronization. Moreover, adding $P_{ref}$-$I_{omag}$ droop control can slightly improve the dynamics of GFM-VSC after grid disturbances, which can be observed by comparing Figs. 18 (b) and (c).

Fig. 19 further illustrates that GFM-VSC with the adaptive fast/slow IVS control can stably ride through the combined large phase angle jump+RoCoF disturbances even if $P_m$ is increased to 1 p.u. under the stiff grid condition. The only difference is that GFM-VSC is already operated with the fast IVS control (FastIVS_en=1) before grid disturbances due to the fact of $I_{omag} > I_{omag\_th\_\min}$.

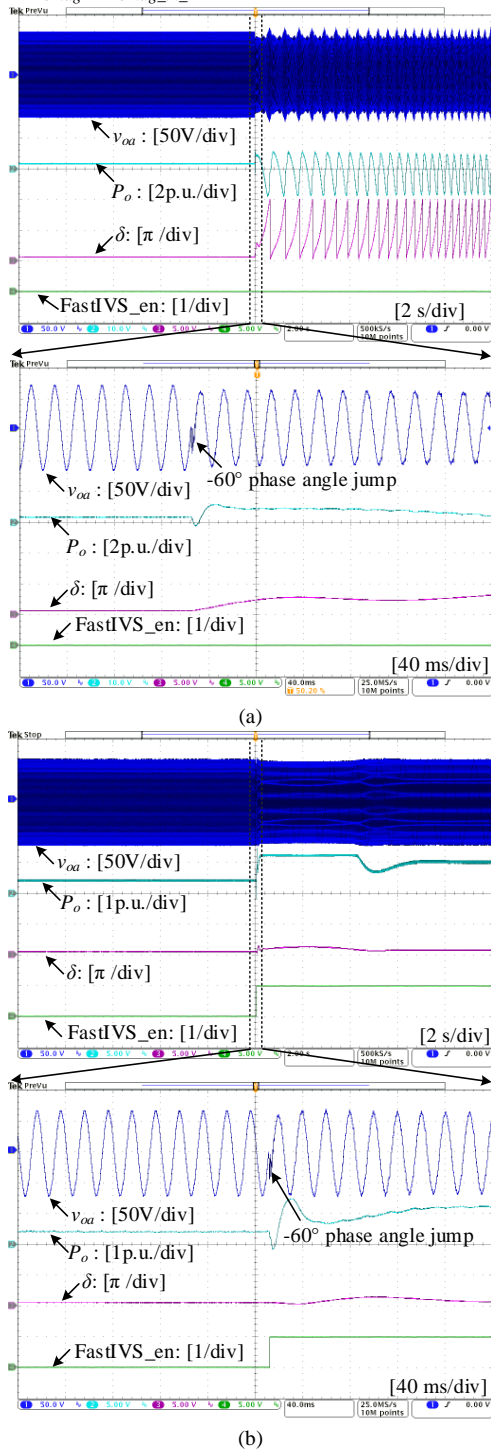

(a)

(b)

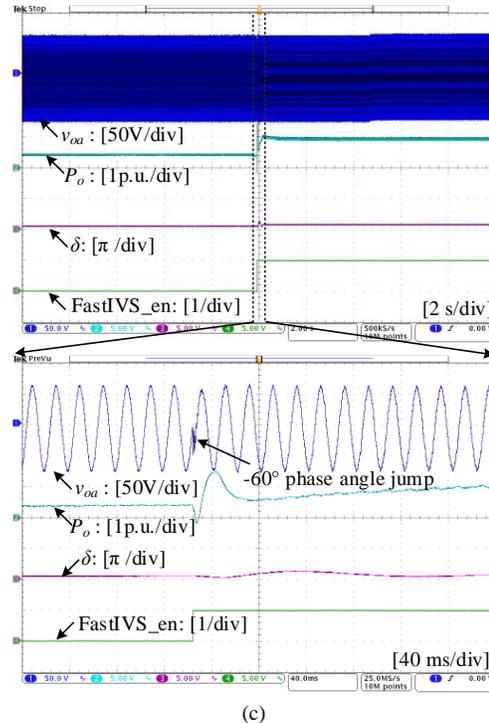

(c)

Fig. 18. Dynamic response of GFM-VSC when subjected to -60° grid phase angle jump ($\theta_g$= -60°), together with -5Hz/s RoCoF that lasts 0.2s. SCR=10, $P_m$=0.4 p.u. (a) Slow IVS control (b) Adaptive fast/slow IVS control without $P_{ref}$-$I_{omag}$ droop control. (c) Adaptive fast/slow IVS control.

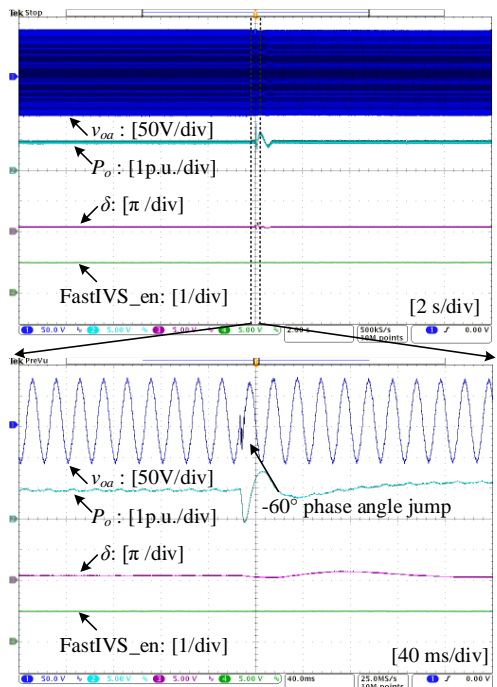

Fig. 19. Dynamic response of GFM-VSC with adaptive fast/slow IVS control when subjected to -60° grid phase angle jump ($\theta_g$= -60°), together with -5Hz/s RoCoF that lasts 0.2s. SCR=10, $P_m$=1 p.u.

Fig. 20 illustrates the dynamic responses of GFM-VSC under the stiff grid condition (SCR=10) considering symmetrical grid faults+phase angle jumps [$V_g$ drops to 0.2 p.u. and lasts for 0.2s, together with -60° phase angle jump ($\theta_g$=

-60°)]. The GFM-VSC is initially operated with $P_m$ =0.7 p.u. It can be seen from Fig. 20(a) that GFM-VSC with slow IVS control (FastIVS_en=0) loses synchronization with the power grid after the fault clearance, which yields one cycle of oscillation where power angle $\delta$ is increased to 360°. Even though GFM-VSC finally re-synchronizes with the power grid, the overall performance is not that satisfactory. In contrast, it can be observed from Figs. 20(b) and (c) that the proposed adaptive control operates GFM-VSC with the fast IVS control (FastIVS_en=1) during grid faults, which enables its direct re-synchronization with the power grid after the fault clearance. Moreover, the GFM-VSC is switched back to the slow IVS control (FastIVS_en=0) when it comes to the steady-state operation after the fault clearance, retaining its GFM capability. Further detailed comparison between Figs. 20(b) and (c) indicates that adding $P_{ref}$-$I_{omag}$ droop control can slightly improve the post-fault dynamics of GFM-VSC in this scenario.

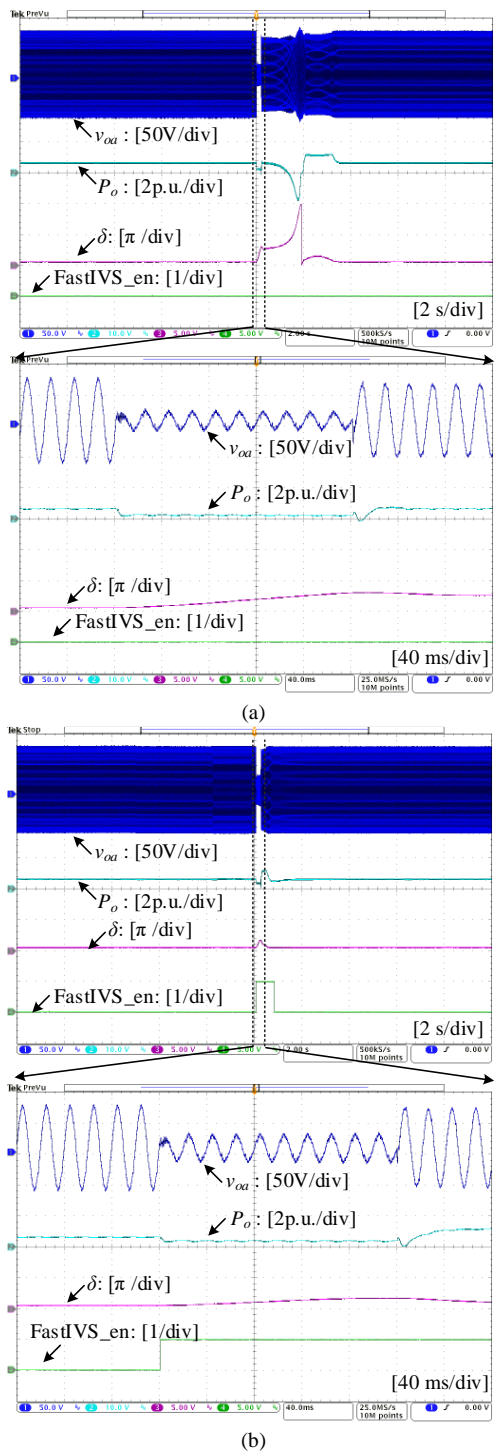

(a)

(b)

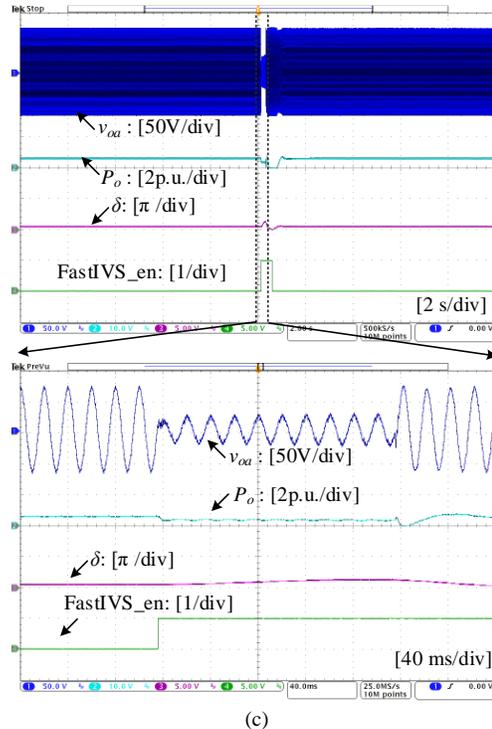

(c)

Fig. 20. Dynamic response of GFM-VSC during symmetrical grid faults ($V_g$ drops to 0.2 p.u. and lasts 0.2s) together with -60° grid phase angle jump ($\theta_g$= -60°). SCR=10, $P_m$=0.7 p.u. (a) Slow IVS control (b) Adaptive fast/slow IVS control without $P_{ref}$-$I_{omag}$ droop control. (c) Adaptive fast/slow IVS control.

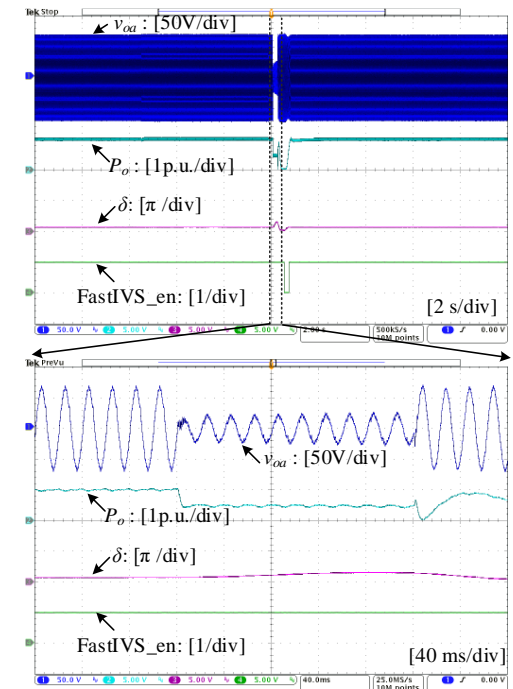

Fig. 21. Dynamic response of GFM-VSC with adaptive fast/slow IVS control during symmetrical grid faults ($V_g$ drops to 0.2 p.u. and lasts 0.2s) together with -60° grid phase angle jump ($\theta_g$= -60°). SCR=10, $P_m$=1 p.u.

Fig. 21 further illustrates that GFM-VSC with the adaptive fast/slow IVS control can stably ride through the combined grid voltage dips+phase angle jump disturbances even if $P_m$ is increased to 1 p.u. under the stiff grid condition. Since the steady-state output current magnitude is larger than $I_{omag\_th\_min}$, GFM-VSC is kept with the fast IVS control (FastIVS_en=1) before and after grid disturbances.

The experimental results given in Figs. 18-21 clearly demonstrate the effectiveness of the proposed adaptive control in guaranteeing the transient stability of GFM-VSC under stiff grid conditions with different kinds of grid disturbances (grid voltage dips, phase angle jump, high RoCoF, and the combination thereof). The proposed $P_{ref}$-$I_{omag}$ droop control enables a slight improvement of the dynamic responses, yet the benefit is not that significant. This is because $V_{omag}$-based $P_{ref}$ adjustment as well as $V_{oq}$ feedforward already perform very well for transient stability enhancement under the stiff grid.

*2) Weak grid (SCR=1.2)*

Fig. 22 illustrates the dynamic responses of GFM-VSC under the weak grid condition (SCR=1.2) considering phase angle jump+RoCoF disturbances [-60° phase angle jump ($\theta_g$= -60°) together with -5Hz/s RoCoF that lasts 0.2s]. The GFM-VSC is initially operated with $P_m$ =0.4 p.u. It can be seen from Fig. 22(a) that GFM-VSC becomes unstable by adopting the slow IVS control. Yet, it also cannot be stabilized with the adaptive control provided the $P_{ref}$ - $I_{omag}$ droop is not used, as

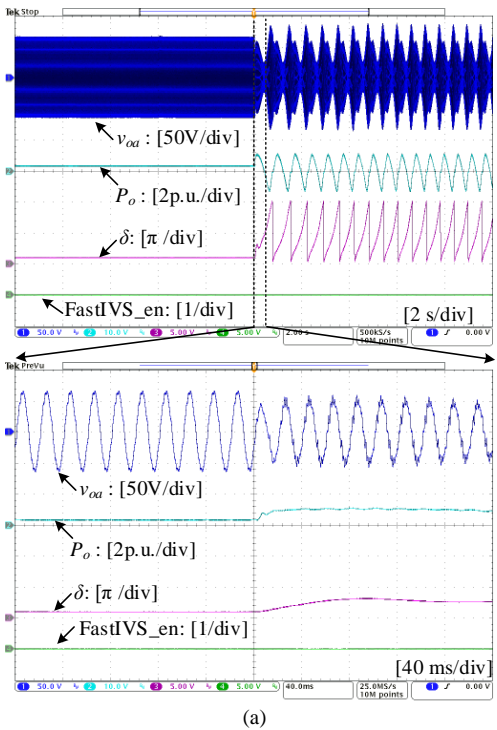

(a)

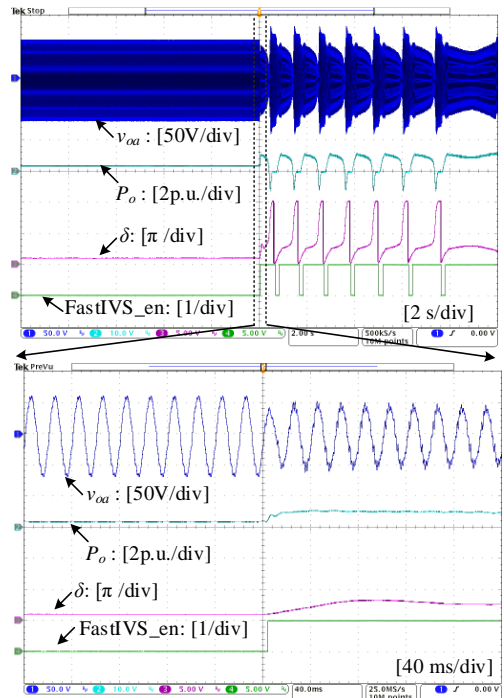

(b)

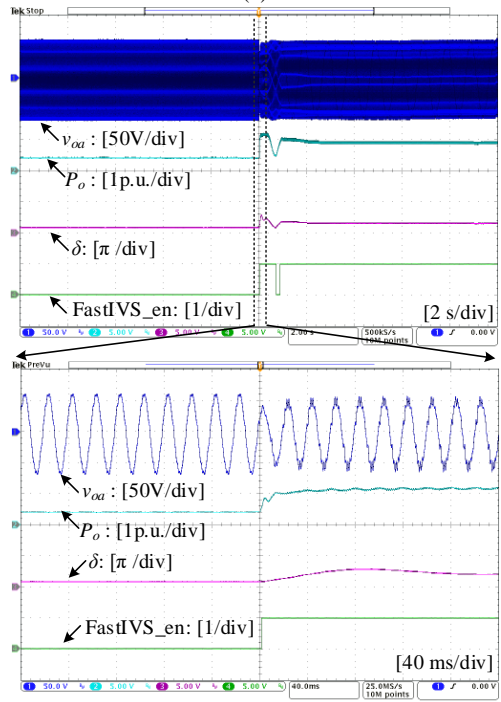

(c)

Fig. 22. Dynamic response of GFM-VSC when subjected to -60° grid phase angle jump ($\theta_g$= -60°), together with -5Hz/s RoCoF that lasts 0.2s. SCR=1.2, $P_m$=0.4 p.u. (a) Slow IVS control (b) Adaptive fast/slow IVS control without $P_{ref}$-$I_{omag}$ droop control. (c) Adaptive fast/slow IVS control.

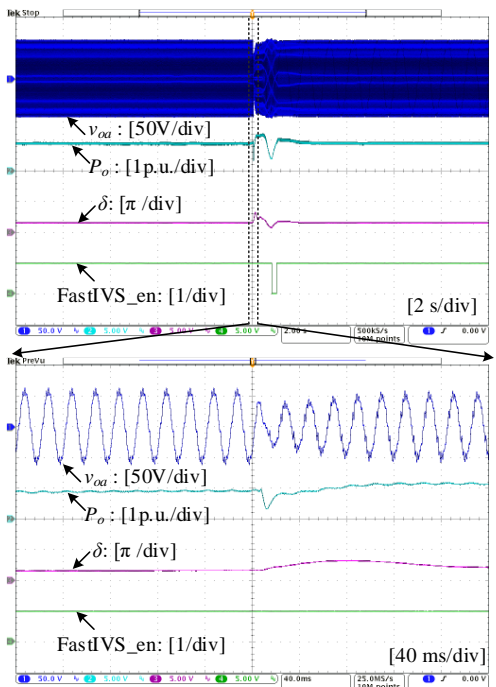

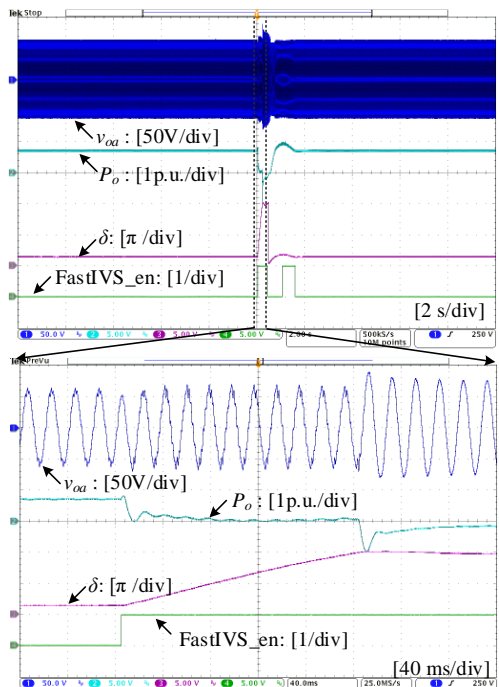

Fig. 23. Dynamic response of GFM-VSC with adaptive fast/slow IVS control when subjected to -60° grid phase angle jump ($\theta_g$= -60°), together with -5Hz/s RoCoF that lasts 0.2s. SCR=1.2, $P_m$=1 p.u.

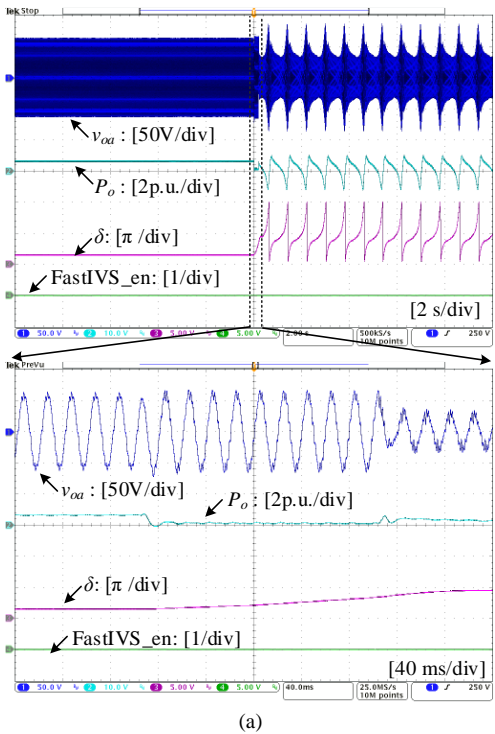

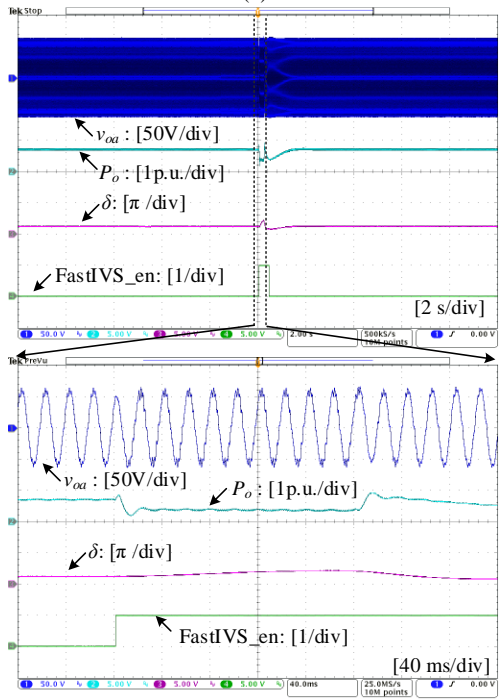

(a)

(b)

(c)

Fig. 24. Dynamic response of GFM-VSC during symmetrical grid faults ($V_g$ drops to 0.2 p.u. and lasts 0.2s) together with -60° grid phase angle jump ($\theta_g$= -60°). SCR=1.2, $P_m$=0.7 p.u. (a) Slow IVS control (b) Adaptive fast/slow IVS control without $P_{ref}$-$I_{omag}$ droop control. (c) Adaptive fast/slow IVS control.

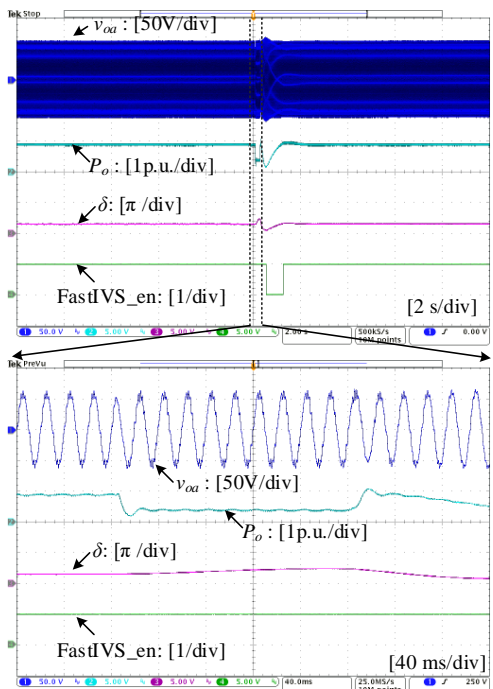

Fig. 25. Dynamic response of GFM-VSC with adaptive fast/slow IVS control during symmetrical grid faults ($V_g$ drops to 0.2 p.u. and lasts 0.2s) together with -60° grid phase angle jump ($\theta_g$= -60°). SCR=1.2, $P_m$=1 p.u.

shown in Fig. 22(b). This is because $V_{omag}$-based $P_{ref}$ adjustment and $V_{oq}$ feedforward become less effective for transient stability enhancement under the weak grid, due to the significant deviations between the PCC voltage and $V_g$, as pointed out in Section IV. Therefore, only the adaptive control with $P_{ref}$-$I_{omag}$ droop included can stabilize GFM-VSC in this scenario, as shown in Fig. 22(c).

Fig. 23 further illustrates that GFM-VSC with adaptive fast/slow IVS control can stably ride through the combined large phase angle jump+RoCoF disturbances even if $P_m$ is increased to 1 p.u. under the weak grid condition. The only difference is that GFM-VSC is already operated with fast IVS control (FastIVS_en=1) before grid disturbances due to the fact of $I_{omag} > I_{omag\_th\_min}$.

Fig. 24 illustrates the dynamic responses of GFM-VSC under the weak grid condition (SCR=1.2) considering symmetrical grid faults+phase angle jumps [$V_g$ drops to 0.2 p.u. and lasts for 0.2s, together with -60° phase angle jump ($\theta_g$= -60°)]. The GFM-VSC is initially operated with $P_m$ =0.7 p.u. It can be seen from Fig. 24(a) that GFM-VSC becomes unstable by adopting the slow IVS control (FastIVS_en=0). Yet, by adopting the adaptive control without the $P_{ref}$ - $I_{omag}$ droop, the GFM-VSC still suffers from one-cycle of oscillation after the fault clearance before it can re-synchronize with the power grid, as shown in Fig. 24(b). This is due to the reduced performance of $V_{omag}$-based $P_{ref}$ adjustment and $V_{oq}$ feedforward under the weak grid. In contrast, by adopting the adaptive control with $P_{ref}$-$I_{omag}$ droop included, the GFM-VSC can successfully ride-through grid faults and directly re-synchronize with the power grid after fault clearance, as shown in Fig. 24(c).

Fig. 25 further illustrates that GFM-VSC with adaptive fast/slow IVS control can stably ride through the combined grid voltage dips+phase angle jump disturbances even if $P_m$ is increased to 1 p.u. under the weak grid condition. Since the steady-state output current magnitude is larger than $I_{omag\_th\_min}$, GFM-VSC is kept with the fast IVS control (FastIVS_en=1) before and after grid disturbances.

Figs. 22-25 clearly demonstrate that the proposed adaptive control can also guarantee the transient stability of GFM-VSC under weak grid conditions with different kinds of grid disturbances (grid voltage dips, phase angle jump, high RoCoF, and the combination thereof). Different from stiff grid scenarios, the proposed $P_{ref}$-$I_{omag}$ droop control is crucial in stabilizing GFM-VSC under the weak grid and should always be included.

*3) Asymmetrical fault-ride through*

Figs. 26-27 illustrate the dynamic response of GFM-VSC under signal-line-to-ground (SLG) fault and double-line-to-ground (DLG) fault considering both stiff and weak grid conditions. It can be seen that with the proposed adaptive fast/slow IVS control, GFM-VSC can stably ride through different kinds of asymmetrical faults regardless of grid strength variations. It is not surprising, as positive-sequence voltage only drops to 0.33 p.u even for the most severe asymmetrical faults, i.e., DLG faults [35]. This represents a better case in terms of transient stability compared with symmetrical faults, in which the remaining positive-sequence voltage could be lower than 0.33 p.u..

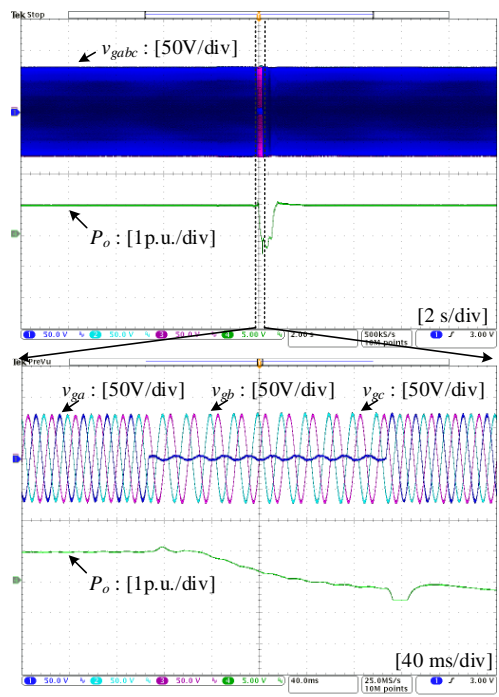

(a)

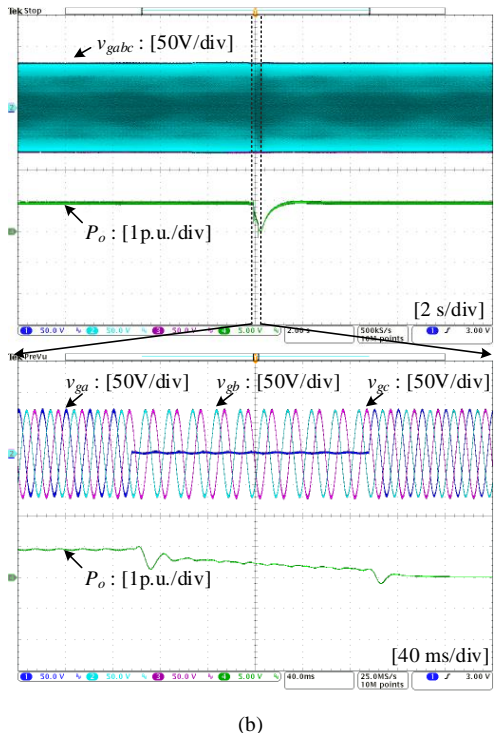

(b)

Fig. 26. Dynamic response of GFM-VSC with adaptive fast/slow IVS control during single-line-to-ground fault. $P_m$=1 p.u. (a) Stiff grid, SCR=10. (b) Weak grid, SCR=1.2.

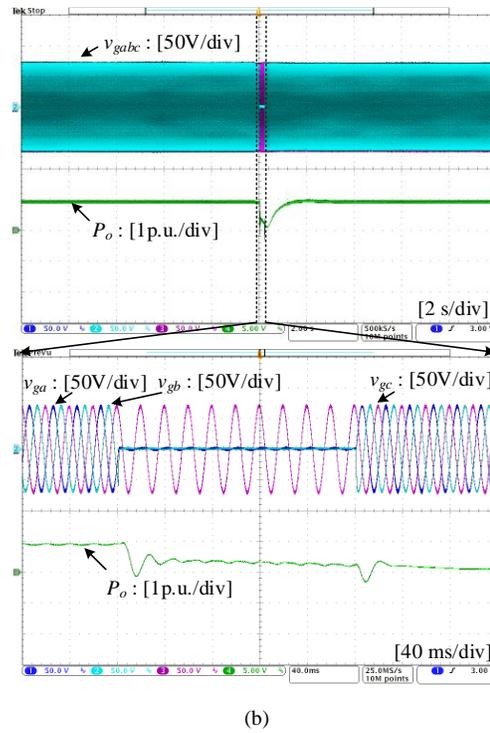

(b)

Fig. 27. Dynamic response of GFM-VSC with adaptive fast/slow IVS control during double-line-to-ground fault. $P_m$=1 p.u. (a) Stiff grid, SCR=10. (b) Weak grid, SCR=1.2.

*4) Compared with other state-of-the-art method*

To further demonstrate the advantage of the proposed adaptive fast/slow IVS control, experimental tests are carried out by comparing the proposed method with the control scheme in [18] that enhances system damping by feedforwarding the frequency in the RPC loop, with which, a good comprise between transient stability and GFM capability can also be achieved. Yet, this method becomes ineffective if GFM-VSC is operated under the current limiting mode where the RPC loop is naturally bypassed. This is demonstrated by experimental results given in Fig. 28(a), where the LOS of GFM-VSC can be observed under large grid disturbances with current limit triggered, even if the method in [18] is used. In contrast, the proposed adaptive fast/slow IVS control allows GFM-VSC to stably ride through large grid disturbances under the current limiting mode, as shown in Fig. 28(b).

## VI. CONCLUSION

With clear physical insights on different IVS dynamic requirements of GFM-VSC for enhancing the transient stability and GFM capability, an adaptive fast/slow IVS control has been proposed in this work. The method not only guarantees the transient stability of GFM-VSC under different grid strength and different types of grid disturbances (grid voltage dips, phase angle jump, high RoCoF, and the combination thereof), but also retains the GFM capability of VSC when there are sufficient stability and current margins. The effectiveness of the proposed method is verified by experimental tests. The efficacy of the proposed adaptive fast/slow IVS control under the large power network with multiple GFM-VSCs will be analyzed in future work.

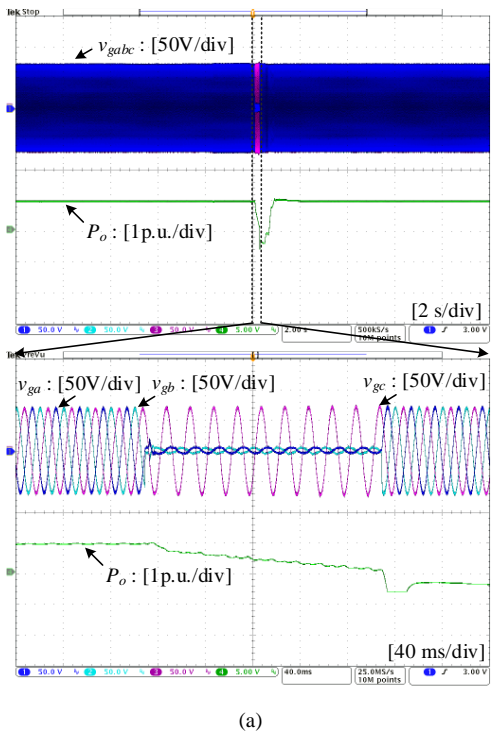

(a)

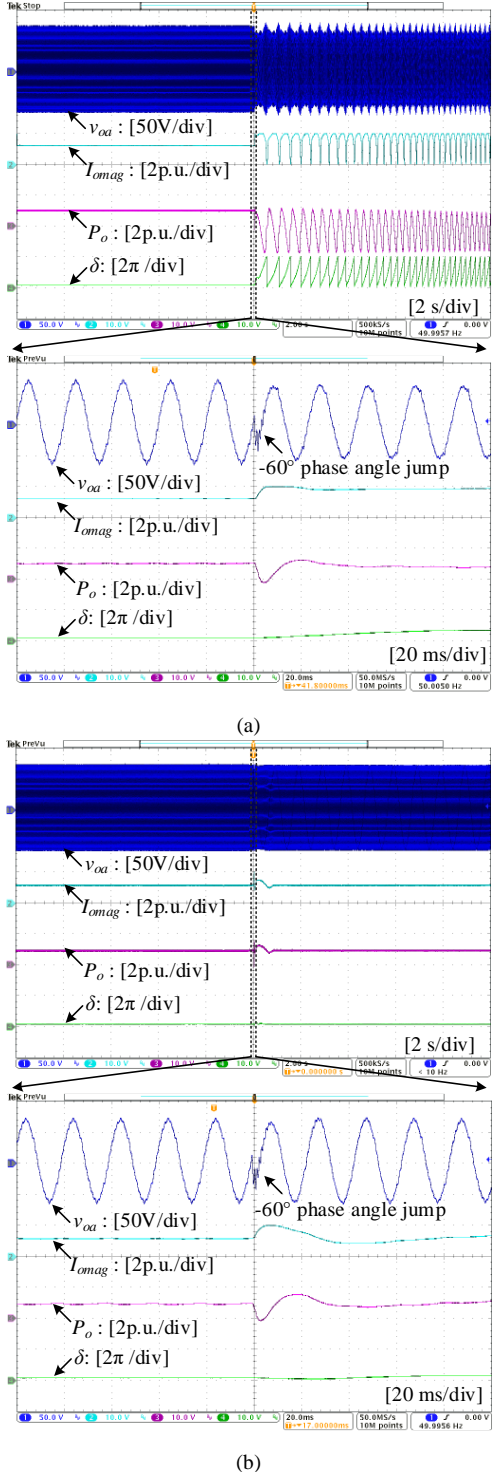

(a)

(b)

Fig. 28. Dynamic response of GFM-VSC when subjected to -60° grid phase angle jump ($\theta_g$= -60°), together with -5Hz/s RoCoF that lasts 0.2s. SCR=10, $P_m$=1 p.u. (a) Existing method in [18], unstable. (b) Proposed adaptive fast/slow IVS control, stable.

## APPENDIX

By considering the dynamics of the APC, there are two rotating dq frames of GFM-VSC [5]: one is the *controller-dq* frame that is defined by the output phase angle of the APC, and another is the *system-dq* frame that is aligned with phase angle of the grid voltage. For clarity, superscript $s$ represents the variables in the *system-dq* frame, while the superscript $c$ denotes the variables in the *controller-dq* frame.

Define $\delta$ as the phase angle difference between the *controller-dq* frame and the *system-dq* frame, i.e., $\delta = \theta_{ref} - \theta_g$. Then the relationships between the state variables in the *controller-dq* frame and the *system-dq* frame are given by

$$x^c_{dq} = e^{-j\delta} x^s_{dq}. \tag{A.1}$$

The small-signal representation of (A.1) is expressed as

$$\hat{x}^c_{dq} = e^{-j\delta_0} \hat{x}^s_{dq} - jX^s_{dq0} e^{-j\delta_0} \hat{\delta} \Leftrightarrow$$

$$\begin{bmatrix} \hat{x}^c_d \\ \hat{x}^c_q \end{bmatrix} = \begin{bmatrix} \cos\delta_0 & \sin\delta_0 \\ -\sin\delta_0 & \cos\delta_0 \end{bmatrix} \begin{bmatrix} \hat{x}^s_d \\ \hat{x}^s_q \end{bmatrix}$$
$$+ \hat{\delta} \begin{bmatrix} \cos\delta_0 & \sin\delta_0 \\ -\sin\delta_0 & \cos\delta_0 \end{bmatrix} \begin{bmatrix} X^s_{q0} \\ -X^s_{d0} \end{bmatrix}. \tag{A.2}$$

It should also be noted that the output of SLVM control, i.e., $E_{ref}$ in Fig. 1, will be saturated during grid disturbances as $V_{omag} \neq V_{ref}$, which naturally bypasses the RPC loop and the SLVM control [32].

### 1) *Small-signal modelling of the power plant and adaptive virtual impedance control*

This part has been addressed in authors' previous publication [32], and hence, only final results will be given. Readers are referred to [32] for more details

The output current in *system-dq* frame can be expressed as

$$\begin{bmatrix} \hat{i}^s_{od} \\ \hat{i}^s_{oq} \end{bmatrix} = \underbrace{\left(\mathbf{Z}_L + \mathbf{Z}_{v\_totdq}\right)^{-1} \left(\mathbf{Z}_{v\_totdq} \begin{bmatrix} -I^s_{oq0} \\ I^s_{od0} \end{bmatrix} + \begin{bmatrix} -V^s_{invq0} \\ V^s_{invd0} \end{bmatrix}\right)}_{G_{i\delta}} \hat{\delta}$$
$$\triangleq \begin{bmatrix} m_1 \\ m_2 \end{bmatrix} \hat{\delta} \tag{A.3}$$

The dynamics of the power plant can be derived as

$$G_{P\delta} = \frac{\hat{p}_o}{\hat{\delta}} = \frac{3}{2} \left( \begin{bmatrix} V^s_{od0} & V^s_{oq0} \end{bmatrix} + \begin{bmatrix} I^s_{od0} & I^s_{oq0} \end{bmatrix} \mathbf{Z}_{L_g} \right) G_{i\delta}. \tag{A.4}$$

### 2) *Small-signal modelling of $V_{oq}$ feedforward*

Based on (A.3) and circuit dynamics of GFM-VSC, $\hat{v}^s_{odq}$ can be expressed as

$$\begin{bmatrix} \hat{v}^s_{od} \\ \hat{v}^s_{oq} \end{bmatrix} = \mathbf{Z}_{L_g} \begin{bmatrix} \hat{i}^s_{od} \\ \hat{i}^s_{oq} \end{bmatrix} = \begin{bmatrix} sm_1 L_g & -m_2 X_g \\ m_1 X_g & sm_2 L_g \end{bmatrix} \hat{\delta} \tag{A.5}$$

Transform (A.5) to *controller-dq* frame based on (A.2), which yields

$$\hat{v}^c_{od} = \underbrace{\left(n_1 \cos\delta_0 + n_2 \sin\delta_0 + V^c_{oq0}\right)}_{G_{vd\delta}} \hat{\delta}$$
$$\hat{v}^c_{oq} = \underbrace{\left(-n_1 \sin\delta_0 + n_2 \cos\delta_0 - V^c_{od0}\right)}_{G_{vq\delta}} \hat{\delta} \tag{A.6}$$

Based on Fig. 12, the dynamics of the $V_{oq}$ feedforward is thus expressed as

$$\Delta\hat{\omega}_q = K_{pvq} \hat{v}^c_{oq} = K_{pvq} G_{vq\delta} \hat{\delta} \tag{A.7}$$

### 3) Small-signal modelling of $V_{omag}$-based $P_{ref}$ adjustment

The output voltage magnitude is calculated as

$$V_{omag} = \sqrt{\left(v_{od}^c\right)^2 + \left(v_{oq}^c\right)^2} \tag{A.8}$$

The small-signal representation of (A.8) is given by:

$$\hat{v}_{omag} = \frac{\partial V_{omag}}{\partial v_{od}^c}\bigg|_{v_{od}^c = V_{od0}^c} \cdot \hat{v}_{od}^c + \frac{\partial V_{omag}}{\partial v_{oq}^c}\bigg|_{v_{oq}^c = V_{oq0}^c} \cdot \hat{v}_{oq}^c$$
$$= \frac{V_{od0}^c \hat{v}_{od}^c + V_{oq0}^c \hat{v}_{oq}^c}{\sqrt{\left(V_{od0}^c\right)^2 + \left(V_{oq0}^c\right)^2}} = \frac{V_{od0}^c \hat{v}_{od}^c + V_{oq0}^c \hat{v}_{oq}^c}{V_{omag0}} \tag{A.9}$$

It is known from Fig. 12 that by using the $V_{omag}$-based $P_{ref}$ adjustment, the modified $P_{ref}$ is expressed as

$$P_{ref} = V_{omag}\left(P_m - D\Delta\omega\right) \tag{A.10}$$

The small-signal representation of (A.10) can be derived as (considering $\Delta\omega_0 = 0$ in the steady-state)

$$\hat{p}_{ref} = P_m \hat{v}_{omag} - D\Delta\omega_0 \hat{v}_{omag} - DV_{omag0}\Delta\hat{\omega}$$
$$= \frac{P_m}{V_{omag0}}\left(V_{od0}^c \hat{v}_{od}^c + V_{oq0}^c \hat{v}_{oq}^c\right) - DV_{omag0}\Delta\hat{\omega} \tag{A.11}$$
$$= k_{v1}\hat{v}_{od}^c + k_{v2}\hat{v}_{oq}^c - DV_{omag0}\Delta\hat{\omega}$$

Substituting (A.6) into (A.11), which yields

$$\hat{p}_{ref} = k_{v1}\hat{v}_{od}^c + k_{v2}\hat{v}_{oq}^c - D\Delta\hat{\omega}$$
$$= \left(k_{v1}G_{vd\delta} + k_{v2}G_{vq\delta}\right)\hat{\delta} - DV_{omag0}\Delta\hat{\omega} \tag{A.12}$$

### 4) Small-signal modelling of $P_{ref}$-$I_{omag}$ droop control

The output current magnitude is calculated as

$$I_{omag} = \sqrt{\left(i_{od}^c\right)^2 + \left(i_{oq}^c\right)^2} \tag{A.13}$$

The small-signal representation of (A.13) is expressed as

$$\hat{i}_{omag} = \frac{\partial I_{omag}}{\partial i_{od}^c}\bigg|_{i_{od}^c = I_{od0}^c} \cdot \hat{i}_{od}^c + \frac{\partial I_{omag}}{\partial i_{oq}^c}\bigg|_{i_{oq}^c = I_{oq0}^c} \cdot \hat{i}_{oq}^c$$
$$= \frac{I_{od0}^c \hat{i}_{od}^c + I_{oq0}^c \hat{i}_{oq}^c}{\sqrt{\left(I_{od0}^c\right)^2 + \left(I_{oq0}^c\right)^2}} = \frac{I_{od0}^c \hat{i}_{od}^c + I_{oq0}^c \hat{i}_{oq}^c}{I_{omag0}}. \tag{A.14}$$
$$= k_{i1}\hat{i}_{od}^c + k_{i2}\hat{i}_{oq}^c$$

Based on (A.2) and (A.3), $\hat{i}_{od}^c$ and $\hat{i}_{oq}^c$ can be expressed as

$$\hat{i}_{od}^c = \underbrace{\left(m_1\cos\delta_0 + m_2\sin\delta_0 + I_{oq0}^c\right)}_{G_{id\delta}}\hat{\delta}$$
$$\hat{i}_{oq}^c = \underbrace{\left(-m_1\sin\delta_0 + m_2\cos\delta_0 - I_{od0}^c\right)}_{G_{iq\delta}}\hat{\delta} \tag{A.15}$$

It is known from Fig. 12 that by using the $P_{ref}$-$I_{omag}$ droop control, the modified $P_{ref}$ is expressed as

$$P_{ref1} = P_{ref} - n\left(I_{omag} - I_{Pth}\right). \tag{A.16}$$

Based on (A.14) and (A.15), the small-signal representation of $P_{ref}$-$I_{omag}$ droop control can be expressed as

$$\hat{p}_{ref1} = \hat{p}_{ref} - n\hat{i}_{omag}$$
$$= \hat{p}_{ref} - n\left(k_{i1}\hat{i}_{od}^c + k_{i2}\hat{i}_{oq}^c\right) \tag{A.17}$$
$$= \hat{p}_{ref} - n\left(k_{i1}G_{id\delta} + k_{i2}G_{iq\delta}\right)\hat{\delta}$$

### 5) Complete small-signal model

The complete small-signal model of GFM-VSC with fast IVS control activated can be derived by combining (A.4), (A.7), (A.12) and (A.17), whose block diagram of given in Fig. A1, based on which, the loop gain can be calculated as

$$T = \frac{G_{P\delta}\left(2HK_p s + 1\right)}{\left\{\begin{array}{l}\left(s - K_{pvq}G_{vq\delta}\right)\left[2Hs + DV_{omag0}\left(2HK_p s + 1\right)\right] \\ + \left(2HK_p s + 1\right)(B - A)\end{array}\right\}} \tag{A.18}$$

where

$$A = k_{v1}G_{vd\delta} + k_{v2}G_{vq\delta}. \tag{A.19}$$
$$B = n\left(k_{i1}G_{id\delta} + k_{i2}G_{iq\delta}\right). \tag{A.20}$$

The closed-loop transfer function is thus calculated as

$$\Phi = \frac{T}{1+T}. \tag{A.21}$$

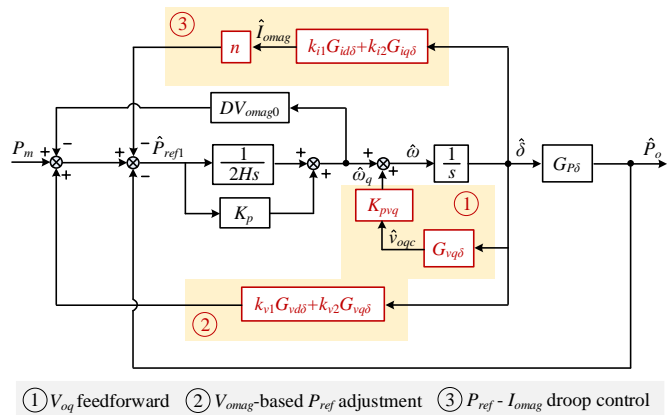

① $V_{oq}$ feedforward  ② $V_{omag}$-based $P_{ref}$ adjustment  ③ $P_{ref}$ - $I_{omag}$ droop control

Fig. A1. Small-signal model of GFM-VSC with fast IVS control activated.

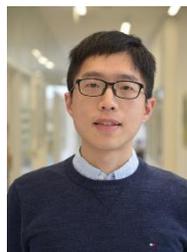

**Heng Wu** (S'17-M' 20) received B.S. and M.S. degrees in electrical engineering from Nanjing University of Aeronautics and Astronautics (NUAA), Nanjing, China, in 2012 and 2015, respectively, and the Ph.D. degree in electrical engineering from Aalborg University, Aalborg, Denmark, in 2020. He is now an Assistant Professor and Leader of Electronic Power Grid (eGRID) Research Group with AAU Energy, Aalborg University.

From 2015 to 2017, He was an Electrical Engineer with NR Electric Co., Ltd, Nanjing, China. He was a guest researcher with Ørsted Wind Power, Fredericia, Denmark, in 2018, and Bundeswehr University Munich, Germany, in 2019. He was a Postdoctoral researcher with Aalborg University, Aalborg, Denmark, from 2020 to 2021. He is the Chairman of IEEE Task Force on Frequency-domain Modeling and Dynamic Analysis of HVDC and FACTS, the subgroup leader of Cigre working group B4/C4.93, the member of GB grid forming best practice expert group formed by national grid ESO, UK, and the member of the technical committee (TC) of European Academy of Wind Energy (EAWE). His research interests include the modelling and stability analysis of the power electronic based power systems. He is identified as world's top 2% scientist by Stanford University from 2019. He received the 2019 Outstanding Reviewer Award of the IEEE

TRANSACTIONS ON POWER ELECTRONICS and the 2021 Star Reviewer Award of the IEEE JOURNAL of EMERGING AND SELECTED TOPICS IN POWER ELECTRONICS

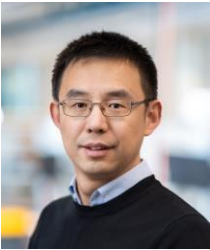

**Xiongfei Wang** (Fellow, IEEE) received the B.S. degree from Yanshan University, China, in 2006, the M.S. degree from Harbin Institute of Technology, China, in 2008, both in electrical engineering, and the Ph.D. degree in energy technology from Aalborg University, Denmark, in 2013.

From 2009 to 2022, he was with Aalborg University where he became an Assistant Professor in 2014, an Associate Professor in 2016, a Professor and Leader of Electronic Power Grid (eGRID) Research Group in 2018. From 2022, he has been a Professor with KTH Royal Institute of Technology, Stockholm, Sweden, and a part-time Professor with Aalborg University, Denmark. His research interests include modeling and control of power electronic converters and systems, stability and power quality of power-electronics-dominated power systems, and high-power converters.

Dr. Wang currently serves as Executive Editor (Editor in Chief) for the IEEE Transactions on Power Electronics Letters and as Associate Editor for the IEEE Journal of Emerging and Selected Topics in Power Electronics. He received ten IEEE Prize Paper Awards, the 2016 AAU Talent for Future Research Leaders, the 2018 Richard M. Bass Outstanding Young Power Electronics Engineer Award, the 2019 IEEE PELS Sustainable Energy Systems Technical Achievement Award, the Clarivate Highly Cited Researcher during 2019-2021, and the 2022 Isao Takahashi Power Electronics Award.